\documentclass[ALICE,manyauthors]{cernphprep}
\usepackage[comma,square,numbers,sort&compress]{natbib}
\usepackage{hyperref}
\usepackage{lineno}
\usepackage{xspace}
\usepackage{cleveref}
\usepackage[T1]{fontenc}
\usepackage{orcidlink}

\begin{document}
%%%%%%%%%%%%%%%%%%%%%%%%%%%%%%%%%%%%%%%%%%%%%%%%%%
% These are some new commands that may be useful 
% for paper writing in general. If other newcommands
% are needed for your specific paper, please feel 
% free to add here. 
%
% The currently available commands are organized in: 
% 1) Systems
% 2) Quantities
% 3) Energies and units
% 4) Detectors
% 5) particle species 
%%%%%%%%%%%%%%%%%%%%%%%%%%%%%%%%%%%%%%%%%%%%%%%%%%

% 1) SYSTEMS 
\newcommand{\pp}           {pp\xspace}
\newcommand{\ppbar}        {\mbox{$\mathrm {p\overline{p}}$}\xspace}
\newcommand{\XeXe}         {\mbox{Xe--Xe}\xspace}
\newcommand{\PbPb}         {\mbox{Pb--Pb}\xspace}
\newcommand{\pA}           {\mbox{pA}\xspace}
\newcommand{\pPb}          {\mbox{p--Pb}\xspace}
\newcommand{\AuAu}         {\mbox{Au--Au}\xspace}
\newcommand{\dAu}          {\mbox{d--Au}\xspace}

% 2) QUANTITIES 
\newcommand{\s}            {\ensuremath{\sqrt{s}}\xspace}
\newcommand{\snn}          {\ensuremath{\sqrt{s_{\mathrm{NN}}}}\xspace}
\newcommand{\pt}           {\ensuremath{p_{\rm T}}\xspace}
\newcommand{\meanpt}       {$\langle p_{\mathrm{T}}\rangle$\xspace}
\newcommand{\ycms}         {\ensuremath{y_{\rm CMS}}\xspace}
\newcommand{\ylab}         {\ensuremath{y_{\rm lab}}\xspace}
\newcommand{\etarange}[1]  {\mbox{$\left | \eta \right |~<~#1$}}
\newcommand{\yrange}[1]    {\mbox{$\left | y \right |~<~#1$}}
\newcommand{\dndy}         {\ensuremath{\mathrm{d}N_\mathrm{ch}/\mathrm{d}y}\xspace}
\newcommand{\dndeta}       {\ensuremath{\mathrm{d}N_\mathrm{ch}/\mathrm{d}\eta}\xspace}
\newcommand{\avdndeta}     {\ensuremath{\langle\dndeta\rangle}\xspace}
\newcommand{\dNdy}         {\ensuremath{\mathrm{d}N_\mathrm{ch}/\mathrm{d}y}\xspace}
\newcommand{\Npart}        {\ensuremath{N_\mathrm{part}}\xspace}
\newcommand{\Ncoll}        {\ensuremath{N_\mathrm{coll}}\xspace}
\newcommand{\dEdx}         {\ensuremath{\textrm{d}E/\textrm{d}x}\xspace}
\newcommand{\RpPb}         {\ensuremath{R_{\rm pPb}}\xspace}

% 3) ENERGIES, UNITS
\newcommand{\nineH}        {$\sqrt{s}~=~0.9$~Te\kern-.1emV\xspace}
\newcommand{\seven}        {$\sqrt{s}~=~7$~Te\kern-.1emV\xspace}
\newcommand{\twoH}         {$\sqrt{s}~=~0.2$~Te\kern-.1emV\xspace}
\newcommand{\twosevensix}  {$\sqrt{s}~=~2.76$~Te\kern-.1emV\xspace}
\newcommand{\five}         {$\sqrt{s}~=~5.02$~Te\kern-.1emV\xspace}
\newcommand{\twosevensixnn}{$\sqrt{s_{\mathrm{NN}}}~=~2.76$~Te\kern-.1emV\xspace}
\newcommand{\fivenn}       {$\sqrt{s_{\mathrm{NN}}}~=~5.02$~Te\kern-.1emV\xspace}
\newcommand{\LT}           {L{\'e}vy-Tsallis\xspace}
\newcommand{\GeVc}         {Ge\kern-.1emV/$c$\xspace}
\newcommand{\MeVc}         {Me\kern-.1emV/$c$\xspace}
\newcommand{\TeV}          {Te\kern-.1emV\xspace}
\newcommand{\GeV}          {Ge\kern-.1emV\xspace}
\newcommand{\MeV}          {Me\kern-.1emV\xspace}
\newcommand{\GeVmass}      {Ge\kern-.2emV/$c^2$\xspace}
\newcommand{\MeVmass}      {Me\kern-.2emV/$c^2$\xspace}
\newcommand{\lumi}         {\ensuremath{\mathcal{L}}\xspace}

% 4) DETECTORS 
\newcommand{\ITS}          {\rm{ITS}\xspace}
\newcommand{\TOF}          {\rm{TOF}\xspace}
\newcommand{\ZDC}          {\rm{ZDC}\xspace}
\newcommand{\ZDCs}         {\rm{ZDCs}\xspace}
\newcommand{\ZNA}          {\rm{ZNA}\xspace}
\newcommand{\ZNC}          {\rm{ZNC}\xspace}
\newcommand{\SPD}          {\rm{SPD}\xspace}
\newcommand{\SDD}          {\rm{SDD}\xspace}
\newcommand{\SSD}          {\rm{SSD}\xspace}
\newcommand{\TPC}          {\rm{TPC}\xspace}
\newcommand{\TRD}          {\rm{TRD}\xspace}
\newcommand{\VZERO}        {\rm{V0}\xspace}
\newcommand{\VZEROA}       {\rm{V0A}\xspace}
\newcommand{\VZEROC}       {\rm{V0C}\xspace}
\newcommand{\Vdecay} 	   {\ensuremath{V^{0}}\xspace}

% 4) PARTICLE SPECIES 
\newcommand{\ee}           {\ensuremath{e^{+}e^{-}}} 
\newcommand{\pip}          {\ensuremath{\pi^{+}}\xspace}
\newcommand{\pim}          {\ensuremath{\pi^{-}}\xspace}
\newcommand{\kap}          {\ensuremath{\rm{K}^{+}}\xspace}
\newcommand{\kam}          {\ensuremath{\rm{K}^{-}}\xspace}
\newcommand{\pbar}         {\ensuremath{\rm\overline{p}}\xspace}
\newcommand{\kzero}        {\ensuremath{{\rm K}^{0}_{\rm{S}}}\xspace}
\newcommand{\lmb}          {\ensuremath{\Lambda}\xspace}
\newcommand{\almb}         {\ensuremath{\overline{\Lambda}}\xspace}
\newcommand{\Om}           {\ensuremath{\Omega^-}\xspace}
\newcommand{\Mo}           {\ensuremath{\overline{\Omega}^+}\xspace}
\newcommand{\X}            {\ensuremath{\Xi^-}\xspace}
\newcommand{\Ix}           {\ensuremath{\overline{\Xi}^+}\xspace}
\newcommand{\Xis}          {\ensuremath{\Xi^{\pm}}\xspace}
\newcommand{\Oms}          {\ensuremath{\Omega^{\pm}}\xspace}
\newcommand{\degree}       {\ensuremath{^{\rm o}}\xspace}

%%%%%%%%%%%%%%%  Title page %%%%%%%%%%%%%%%%%%%%%%%%
\begin{titlepage}
% the dates below correspond to CERN approval
% please don't touch: EB chairs will take care
\PHyear{2023}       % required, will be obtained from CERN
\PHnumber{187}      % required, will be obtained from CERN
\PHdate{28 August}  % required, will be obtained from CERN
%%%%%%%%%%%%%%%%%%%%%%%%%%%%%%%%%%%%%%%%%%%%%%%%%%%%

%%% Put your own title + short title here:
\title{Skewness and kurtosis of mean transverse momentum fluctuations at the LHC energies}
\ShortTitle{Higher-order $\langle p_\mathrm{T}\rangle$ fluctuations at LHC energies}   % appears on left page headers

%%% Do not change the next lines
\Collaboration{ALICE Collaboration\thanks{See Appendix~\ref{app:collab} for the list of collaboration members}}
\ShortAuthor{ALICE Collaboration} % appears on right page headers, do not change

\begin{abstract}
  The first measurements of skewness and kurtosis of mean transverse momentum ($\langle p_\mathrm{T}\rangle$) fluctuations are reported in Pb--Pb collisions at $\sqrt{s_\mathrm{NN}}$ $=$ 5.02 TeV, Xe--Xe collisions at $\sqrt{s_\mathrm{NN}}$ $=$ 5.44 TeV  and pp collisions at $\sqrt{s} = 5.02$ TeV using the ALICE detector. The measurements are carried out as a function of system size $\langle \mathrm{d}N_\mathrm{ch}/\mathrm{d}\eta\rangle_{|\eta|<0.5}^{1/3}$, using charged particles with transverse momentum ($p_\mathrm{T}$) and pseudorapidity ($\eta$), in the range $0.2 < p_\mathrm{T} < 3.0$ GeV/$\it{c}$ and $|\eta| < 0.8$, respectively.
  In Pb--Pb and Xe--Xe collisions, positive skewness is observed in the fluctuations of $\langle p_\mathrm{T}\rangle$ for all centralities, which is significantly larger than what would be expected in the scenario of independent particle emission. This positive skewness is considered a crucial consequence of the hydrodynamic evolution of the hot and dense nuclear matter created in heavy-ion collisions. 
  Furthermore, similar observations of positive skewness for minimum bias pp collisions are also reported here. Kurtosis of $\langle p_\mathrm{T}\rangle$ fluctuations is found to be in good agreement with the kurtosis of Gaussian distribution, for most central Pb--Pb collisions. Hydrodynamic model calculations with MUSIC using Monte Carlo Glauber initial conditions are able to explain the measurements of both skewness and kurtosis qualitatively from semicentral to central collisions in Pb--Pb system. Color reconnection mechanism in PYTHIA8 model seems to play a pivotal role in capturing the qualitative behavior of the same measurements in pp collisions.

\end{abstract}
\end{titlepage}

\setcounter{page}{2} %please do not remove this line

%%%%%%%%%%%%%%%%%%%%%%%%%%%%%%%%
% begin main text
%%%%%%%%%%%%%%%%%%%%%%%%%%%%%%%%

\section{Introduction}
The properties of the hot and dense nuclear matter created in heavy-ion collisions at relativistic energies can be studied using event-by-event fluctuations of different quantities like multiplicity, net-charge, mean transverse momentum ($\langle p_\text{T}\rangle$), etc.~\cite{Shuryak:1997yj, Stephanov:1999zu, Heiselberg:2000fk, ALICE:2022wpn}. The analysis of event-by-event fluctuations of these variables offers valuable means to probe the dynamical fluctuations that originate from the production of a quark--gluon plasma (QGP)~\cite{Shuryak:1978ij} phase during heavy-ion collisions. 
Fluctuations in the thermodynamic quantity of temperature, which are associated with the phase transition in the quantum chromodynamics (QCD) phase diagram, can manifest themselves in the fluctuations of the $\langle p_\text{T}\rangle$ of the final-state particles~\cite{Stodolsky:1995ds}. A non-monotonic behavior of $\langle p_\text{T}\rangle$ fluctuations as a function of centrality or incident energy was suggested as one of the possible signals of the QGP~\cite{Heiselberg:2000fk}. However, $\langle p_\text{T}\rangle$ fluctuations are also affected by non-thermodynamic variations in the initial geometry of the collision that include fluctuations in the initial size, shape and orientation of the colliding nuclei, and the fluctuating number of nucleons participating in the collision. The $\langle p_\text{T}\rangle$ fluctuations measured at RHIC did not show any beam energy dependence, and the non-monotonic behavior with centrality was also not observed~\cite{STAR:2005vxr}. The previous measurements of $\langle p_\text{T}\rangle$ fluctuations in Pb--Pb collisions at centre-of-mass energy per nucleon pair, $\sqrt{s_\mathrm{NN}}$ $=$ 2.76 TeV in ALICE~\cite{ALICE:2014gvd} suggested a connection of the observed fluctuations of the $\langle p_\text{T}\rangle$ to the fluctuations in the initial state of the collision. Qualitatively, the data obtained from model calculations using the string melting approach in A Multi-Phase Transport (AMPT) model~\cite{Lin:2004en}, where partons rescatter and recombine through a hadronic coalescence scheme, exhibit agreement with the observed results~\cite{ALICE:2014gvd}. The string melting version was introduced in the model~\cite{Zhang:2019utb} to address and account for the effects of flow originating from the entire partonic system in the overlap volume of heavy-ion collisions, as opposed to considering only the contributions from minijet partons in the default version. By incorporating the string melting approach, the model was able to simulate the collective behavior of the partonic system during the early stages of the collision, leading to a more accurate description of the experimental observations of elliptic flow. 
The $\langle p_\text{T}\rangle$ fluctuations were also calculated in the color glass condensate (CGC)~\cite{Gelis:2010nm, Weigert:2005us} formulation, where they have been related to initial spatial fluctuations of glasma flux tube via their coupling to a collective flow field. The comparison of these calculations with data showed a good agreement in the semicentral and central collisions~\cite{Gavin:2011gr}.

The space--time evolution of the QGP phase produced in the relativistic heavy-ion collisions is well described by the relativistic viscous hydrodynamics~\cite{Bhadury:2020ivo, Heinz:2011kt, Romatschke:2009im}. In Ref.~\cite{SkewnessTheoryPRC}, it has been proposed that the skewness of $\langle p_\text{T}\rangle$ fluctuations can serve as an essential probe of the hydrodynamic behavior of the system created in heavy-ion collisions. The $\langle p_\text{T}\rangle$ of the particles emitted at freeze-out was found to be correlated to the initial energy of the system at the beginning of the hydrodynamic evolution, instead of the energy of the system at freeze-out~\cite{Gardim:2020sma}. The fluctuations of $\langle p_\text{T}\rangle$ are therefore found to be related to the fluctuations of initial energy density in an $\it{effective}$ hydrodynamic description~\cite{Gardim:2019xjs}. It was shown that the skewness of $\langle p_\text{T}\rangle$ fluctuations are driven by the skewness of the initial energy density fluctuations, which implies that the $\langle p_\text{T}\rangle$ fluctuations arise from the same collective dynamics in the QGP phase that give rise to anisotropic flow. 
Furthermore, employing initial conditions from the $\rm T_{R}ENTo$ model~\cite{Moreland:2014oya} and evolving them with the \textsc{v-usphydro} viscous hydrodynamic~\cite{Noronha-Hostler:2013gga} simulations predicted positive skewness of $\langle p_\text{T}\rangle$ fluctuations surpassing expectations in independent particle emission scenarios. Additional information regarding these simulations can be found in Ref.~\cite{SkewnessTheoryPRC}.

The fluctuations in $\langle p_\text{T}\rangle$ of charged particles can be influenced by various physical effects, such as collective behavior of the system formed in the collisions, fluctuations in the number of participating nucleons, or the presence of jets and resonance decays. Examining the higher-order terms of $\langle p_\text{T}\rangle$ fluctuations will enable us to delve deeper into the intricate mechanisms underlying the observed fluctuations and acquire valuable insights. In this article, the first experimental study of skewness and kurtosis of $\langle p_\text{T}\rangle$ fluctuations that represent the third- and fourth-order fluctuations of $\langle p_\text{T}\rangle$ are reported at the Large Hadron Collider (LHC) energies. The observables used in this analysis are introduced in Sec.~\ref{sec:Obs}. Here, the multiparticle $p_\text{T}$ correlators used to study the $\langle p_\text{T}\rangle$ fluctuations are defined. A brief description of the subsystems of ALICE detector relevant to this analysis is given in Sec.~\ref{sec:ExpData}. The analysis technique and the method of estimating statistical and systematic uncertainties for the measurements are explained in Sec.~\ref{sec:DataAna}. The skewness and the kurtosis of $\langle p_\text{T}\rangle$ fluctuations as a function of system size and the interpretation of results using theoretical models are presented in Sec.~\ref{sec:res}. The major findings of the analysis are summarized in Sec.~\ref{sec:summary}.

\section{Observables}
\label{sec:Obs}

In this analysis, the fluctuations in event-by-event $\langle p_\text{T}\rangle$ of charged particles are investigated using multiparticle $p_\mathrm{T}$ correlators. The event-by-event $\langle p_\text{T}\rangle$ is defined as
\begin{equation}
  \label{eq:meanpt}
  \langle p_\text{T}\rangle = \frac{\sum_{i=1}^{N_\text{ch}}p_{\text{T},i}}{N_\text{ch}}\,,
\end{equation}
where $p_{\text{T},i}$ is the transverse momentum of the $\it{i}$th particle and $N_\mathrm{ch}$ is the total number of charged particles in the event. Alternatively, one can employ the standard moment method for event-by-event analysis of $\langle p_\text{T}\rangle$ fluctuations. The moment method calculates various order moments of the $\langle p_\text{T}\rangle$ distribution, providing a comprehensive evaluation of the total fluctuation accounting for both the statistical and dynamical (non-statistical) parts. The advantage of employing multiparticle $p_\mathrm{T}$ correlators~\cite{Voloshin:2001ei, Voloshin:2002ku} is that they yield zero values for events with randomly sampled particles, thereby effectively isolating the non-statistical fluctuations of interest.
The expressions for the two-particle $p_\mathrm{T}$ correlator ($\langle \Delta p_{\text{T},i}\Delta p_{\text{T},j}\rangle$) and three-particle $p_\mathrm{T}$ correlator ($\langle \Delta p_{\text{T},i}\Delta p_{\text{T},j}\Delta p_{\text{T},k}\rangle$)~\cite{SkewnessTheoryPRC} are denoted by Eqs.~\ref{eq:simp2} and~\ref{eq:simp3}, respectively, where $Q_n$ is defined as
\begin{align}
Q_n=\sum_{i=1}^{N_\mathrm{ch}}p_{\text{T},i}^{n}\,.
\label{eq:Qn}
\end{align}
Constraints on the indices in summations in the definition of correlators ensure that self-correlations are eliminated. The algebraic expressions of the correlators in terms of $Q_n$'s where $n = 1, 2, ...$ were derived to account for the practical limitations of calculating multiparticle $p_\mathrm{T}$ correlators with nested loops, especially when dealing with events characterized by large multiplicities. Since, the values of $Q_n$ can be obtained with Eq.~\ref{eq:Qn} employing a single loop, it becomes feasible to directly utilize the later expressions with $Q_n$s in Eqs.~\ref{eq:simp2} and~\ref{eq:simp3} for the calculation of two- and three-particle $p_\mathrm{T}$ correlators, respectively. On similar lines, the four-particle $p_\mathrm{T}$ correlator ($\langle \Delta p_{\text{T},i}\Delta p_{\text{T},j}\Delta p_{\text{T},k}\Delta p_{\text{T},l}\rangle$) is derived as shown in Eq.~\ref{eq:simp4}. 
These two-, three-, and four-particle correlators are related to the variance, skewness, and kurtosis of the $\langle p_\text{T}\rangle$ distribution, respectively.

\begin{align}
  \langle \Delta p_{\text{T},i}\Delta p_{\text{T},j}\rangle =  \left<\frac{\sum_{\substack{i, j,\\ i\neq j}}^{N_\mathrm{ch}}(p_{\text{T},i}-\langle\langle p_\text{T}\rangle\rangle)(p_{\text{T},j}-\langle\langle p_\text{T}\rangle\rangle)}{N_\mathrm{ch}(N_\mathrm{ch}-1)}\right>_\mathrm{ev} = \left<\frac{Q_1^2-Q_2}{N_\mathrm{ch}(N_\mathrm{ch}-1)}\right>_\mathrm{ev}-\left<\frac{Q_1}{N_\mathrm{ch}}\right>^{2}_\mathrm{ev}\,,
  \label{eq:simp2}
\end{align}

\begin{align}
  \langle \Delta p_{\text{T},i}\Delta p_{\text{T},j}\Delta p_{\text{T},k}\rangle &= \left<\frac{\sum_{\substack{i, j, k,\\ i\neq j\neq k}}^{N_\mathrm{ch}}(p_{\text{T},i}-\langle\langle p_\text{T}\rangle\rangle)(p_{\text{T},j}-\langle\langle p_\text{T}\rangle\rangle)(p_{\text{T},k}-\langle\langle p_\text{T}\rangle\rangle)}{N_\mathrm{ch}(N_\mathrm{ch}-1)(N_\mathrm{ch}-2)}\right>_\mathrm{ev}\nonumber \\ &=\left<\frac{Q_1^3-3Q_2Q_1+2Q_3}{N_\mathrm{ch}(N_\mathrm{ch}-1)(N_\mathrm{ch}-2)}\right>_\mathrm{ev}-3\left<\frac{Q_1^2-Q_2}{N_\mathrm{ch}(N_\mathrm{ch}-1)}\right>_\mathrm{ev}\left<\frac{Q_1}{N_\mathrm{ch}}\right>_\mathrm{ev}+2\left<\frac{Q_1}{N_\mathrm{ch}}\right>^3_\mathrm{ev}\,,
  \label{eq:simp3}
\end{align} 

\begin{align}
  \langle\Delta p_{\text{T},i}\Delta p_{\text{T},j}\Delta p_{\text{T},k}\Delta p_{\text{T},l}\rangle &= \left<\frac{\sum_{\substack{i, j, k, l,\\ i\neq j\neq k\neq l}}^{N_\mathrm{ch}}(p_{\text{T},i}-\langle\langle p_\text{T}\rangle\rangle)(p_{\text{T},j}-\langle\langle p_\text{T}\rangle\rangle)(p_{\text{T},k}-\langle\langle p_\text{T}\rangle\rangle)(p_{\text{T},l}-\langle\langle p_\text{T}\rangle\rangle)}{N_\mathrm{ch}(N_\mathrm{ch}-1)(N_\mathrm{ch}-2)(N_\mathrm{ch}-3)}\right>_\mathrm{ev} \nonumber\\
  &=\left<\frac{Q_1^4-6Q_4+8Q_1Q_3-6Q_1^2Q_2+3Q_2^2}{N_\mathrm{ch}(N_\mathrm{ch}-1)(N_\mathrm{ch}-2)(N_\mathrm{ch}-3)}\right>_\mathrm{ev}\nonumber\\
  &-4\left<\frac{Q_1^3-3Q_2Q_1+2Q_3}{N_\mathrm{ch}(N_\mathrm{ch}-1)(N_\mathrm{ch}-2)}\right>_\mathrm{ev}\left<\frac{Q_1}{N_\mathrm{ch}}\right>_\mathrm{ev}\nonumber\\
  &+6\left<\frac{Q_1^2-Q_2}{N_\mathrm{ch}(N_\mathrm{ch}-1)}\right>_\mathrm{ev}\left<\frac{Q_1}{N_\mathrm{ch}}\right>_\mathrm{ev}^2-3\left<\frac{Q_1}{N_\mathrm{ch}}\right>_\mathrm{ev}^4\,.
  \label{eq:simp4}
\end{align}

In the above equations, angular brackets $\langle\ldots\rangle_\mathrm{ev}$ denote an average over all events and $\langle\langle p_\text{T}\rangle\rangle=\langle Q_1/N_\mathrm{ch}\rangle_\mathrm{ev}$ is the event-by-event $\langle p_\text{T}\rangle$ averaged over all events. 
The standardized skewness ($\gamma_{\langle p_\text{T}\rangle}$) and the intensive skewness ($\Gamma_{\langle p_\mathrm{T}\rangle}$) are two measures of the skewness of the $\langle p_\text{T}\rangle$ distribution, constructed using the two-particle and three-particle $p_\mathrm{T}$ correlators~\cite{SkewnessTheoryPRC}, given by 
\begin{equation}
\gamma_{\langle p_\mathrm{T}\rangle}=\frac{\langle \Delta p_{\text{T},i}\Delta p_{\text{T},j}\Delta p_{\text{T},k}\rangle}{\langle \Delta p_{\text{T},i}\Delta p_{\text{T},j}\rangle^{3/2}}\,,
  \label{eq:standardized_skewness}
\end{equation}
and
\begin{equation}
\Gamma_{\langle p_\mathrm{T}\rangle}=\frac{\langle \Delta p_{\text{T},i}\Delta p_{\text{T},j}\Delta p_{\text{T},k}\rangle\langle\langle p_\text{T}\rangle\rangle}{\langle \Delta p_{\text{T},i}\Delta p_{\text{T},j}\rangle^{2}}\,.
  \label{eq:intensive_skewness}
\end{equation}

The kurtosis $\kappa_{\langle p_\mathrm{T}\rangle}$ of $\langle p_\text{T}\rangle$ distribution is defined as
\begin{equation}
  \kappa_{\langle p_\mathrm{T}\rangle}=\frac{\langle\Delta p_{\text{T},i}\Delta p_{\text{T},j}\Delta p_{\text{T},k}\Delta p_{\text{T},l}\rangle}{\langle\Delta p_{\text{T},i}\Delta p_{\text{T},j}\rangle^{2}}\quad.
  \label{eq:dynamic_kurtosis}
\end{equation}
The article focuses on analyzing higher order fluctuations of $\langle p_\mathrm{T}\rangle$, particularly the third and fourth order. Three key observables, namely the standardized skewness, intensive skewness, and kurtosis of the $\langle p_\text{T}\rangle$ distribution (defined in Eqs.~\ref{eq:standardized_skewness},~\ref{eq:intensive_skewness}, and~\ref{eq:dynamic_kurtosis}, respectively), are investigated as a function of the system size in Pb--Pb collisions at $\sqrt{s_\mathrm{NN}}$ $=$ 5.02 TeV, Xe--Xe collisions at $\sqrt{s_\mathrm{NN}}$ = 5.44 TeV, and pp collisions at $\sqrt{s}$ $=$ 5.02 TeV. 

\section{Experimental setup}
\label{sec:ExpData}

\begin{figure}[tb]
    \begin{center}
    \includegraphics[width = 0.5\textwidth]{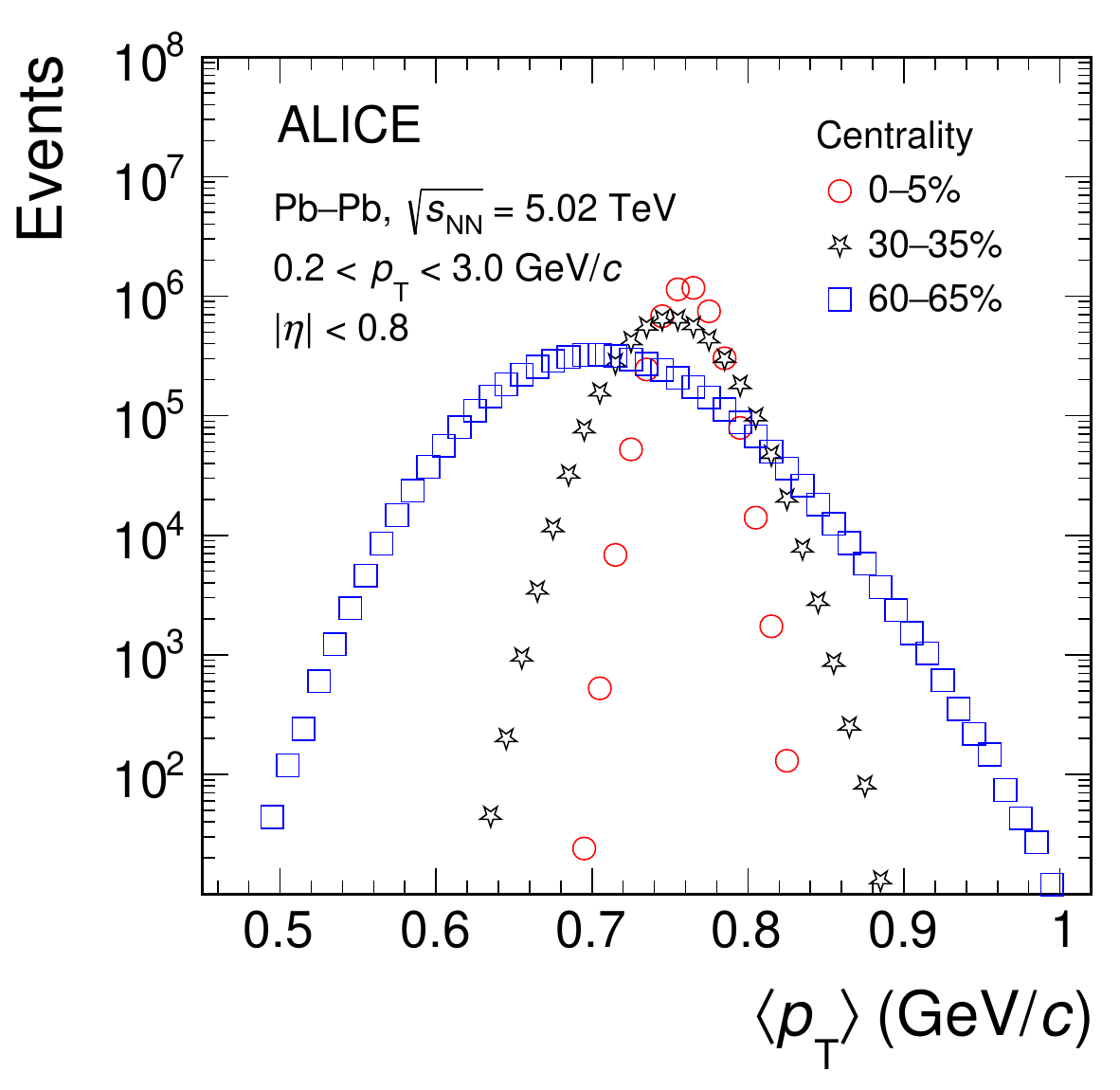}
    \end{center}
    \caption{Efficiency uncorrected mean transverse momentum distributions for different centrality classes in Pb--Pb collisions at $\sqrt{\it{s}_\mathrm{NN}}$ $=$ 5.02 TeV.}
    \label{fig:meanpt_dist}
\end{figure}
The measurements reported in the article are obtained using the data recorded by the ALICE detector at the LHC. A detailed description of the ALICE detector and its performance can be found in Refs.~\cite{ALICE:2008ngc, ALICE:2014sbx}. The primary sub-detectors relevant to this analysis are the Time Projection Chamber (TPC)~\cite{AliceTPC}, the Inner Tracking System (ITS)~\cite{ALICE:2014sbx}, and the V0 detector~\cite{ALICE:V0}. The TPC and ITS are used for tracking and reconstructing the primary vertex, while the V0 detector is used for triggering and the default centrality estimation. The V0 detector consists of two scintillator arrays, V0A and V0C, located on both sides of the interaction point, covering the pseudorapidity ($\eta$) intervals  $2.8 < \eta < 5.1$ and $-3.7 < \eta  < -1.7$, respectively~\cite{ALICE:V0}. The data analyzed here are obtained from Pb--Pb collisions at $\sqrt{s_\mathrm{NN}}$ $=$ 5.02 TeV, Xe--Xe collisions at  $\sqrt{s_\mathrm{NN}}$ = 5.44 TeV, and pp collisions at $\sqrt{s}$ $=$ 5.02 TeV. These datasets were recorded in 2015 (pp and Pb--Pb) and 2017 (Xe--Xe) during LHC Run 2. A minimum bias (MB) trigger condition is applied to select collision events that requires at least one hit in both the V0A and the V0C detectors. Events that pass the MB trigger criteria and have a reconstructed primary vertex position within 10 cm along the beam axis to the nominal interaction point are selected in the analysis. Using information from multiple detectors as described in Ref.~\cite{ ALICE:2014sbx}, events with more than one reconstructed primary interaction vertex (pile-up events) are rejected. In total, 84 million Pb--Pb collisions, 1.2 million Xe--Xe collisions, and 95 million pp collisions pass the above mentioned criteria and are used for the analysis.

The charged-particle tracks reconstructed using the ITS and the TPC in the ALICE central barrel are selected in the kinematic range $0.2 < p_\mathrm{T} < 3.0$ $\mathrm{GeV}/c$ and $|\eta| < 0.8$, where $p_\mathrm{T}$ is the track momentum in the plane transverse to the beam axis. The accepted tracks exhibit approximately uniform azimuthal acceptance in this region. 
The tracks that have at least 70 out of a maximum possible 159 reconstructed space points in the TPC, and at least one hit in the two innermost layers of the ITS (ITS has in total six layers) are selected. This selection assures a resolution better than 300 $\mu$m~\cite{ALICE:2014sbx} on the distance-of-closest-approach (DCA) to the primary vertex in the plane perpendicular ($\mathrm{DCA}_{xy}$) and parallel ($\mathrm{DCA}_{z}$) to the beam axis for the selected tracks. In order to suppress the contamination from secondary particles, the DCA of the tracks to the primary vertex must be within 1 cm in the longitudinal direction and 0.1 cm in the transverse plane. Moreover, the $\chi^{2}$ per degree of freedom of the track fit to the space points in the TPC and the ITS are required to be less than 4 (2.5 for Pb--Pb) and 36, respectively.

\section{Data analysis}
\label{sec:DataAna}

\begin{figure}[tb]
    \begin{center}
    \includegraphics[width = 0.95\textwidth]{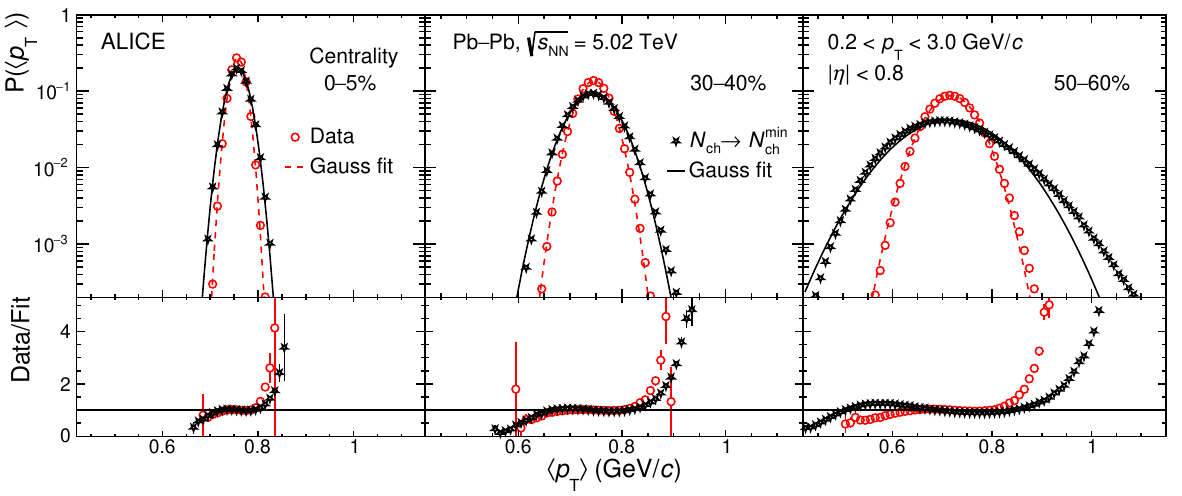}
    \end{center}
    \caption{Comparison of event-by-event mean transverse momentum $\langle p_\mathrm{T}\rangle$ distribution from actual analysis and modified analysis in which $N_\mathrm{ch}$ in each event is fixed to $N_\mathrm{ch}^\mathrm{min}$ ($N_\mathrm{ch}^\mathrm{min}$ is the minimum number of charged particles in a given centrality class) and $N_\mathrm{ch}^\mathrm{min}$ number of particles are selected in each event randomly for the 0--5\% (left), 30--40\% (middle), and 50--60\% (right) centrality classes.}
    \label{fig:meanpt_dist_2}
\end{figure}

The mean transverse momentum $\langle p_\mathrm{T}\rangle$ distributions obtained for three different centrality classes in Pb--Pb collisions at $\sqrt{s_\mathrm{NN}}$ $=$ 5.02 TeV are shown in Fig.~\ref{fig:meanpt_dist}. The distributions are not corrected for detector inefficiencies. The centrality classes are formed by splitting the events based on the measured amplitude distribution in the V0A and V0C detectors as described in Refs.~\cite{ALICE:2018tvk, ALICE:2018yvr}. The mean of the $\langle p_\mathrm{T}\rangle$ distributions are found to increase whereas the width of the distribution decreases from peripheral to central collisions. The larger width of the distribution for peripheral (lower multiplicity) collisions indicates larger fluctuations compared to the central (higher multiplicity) collisions. Similar behavior is observed in Xe--Xe collisions at $\sqrt{s_\mathrm{NN}}$ $=$ 5.44 TeV and pp collisions at $\sqrt{s}$ $=$ 5.02 TeV. 

Before starting the analysis with the $\langle p_\mathrm{T}\rangle$ correlators presented in Sec.~\ref{sec:Obs}, it is important to demonstrate that these fluctuations are not a trivial consequence of fluctuating $N_\mathrm{ch}$ in a given centrality class. The definition of $\langle p_\mathrm{T}\rangle$ in Eq.~\ref{eq:meanpt} clearly shows that the varying multiplicity in different events of the same centrality class can affect the $\langle p_\mathrm{T}\rangle$ distributions and consequently its fluctuations. To examine whether these fluctuations persist after removing the stochastic effects of the multiplicity, a check is performed. For each centrality class, the minimum number of tracks per event in the kinematic acceptance of the analysis, $N_\mathrm{ch}^\mathrm{min}$ is determined. In each event of the centrality class, the $N_\mathrm{ch}^\mathrm{min}$ number of tracks are then selected randomly to calculate the genuine $\langle p_\mathrm{T}\rangle$ distribution free from biases of multiplicity fluctuations. The left panel of Fig.~\ref{fig:meanpt_dist_2} shows the $\langle p_\mathrm{T}\rangle$ distribution for the 0--5\% central events in Pb--Pb collisions at $\sqrt{s_\mathrm{NN}}$ $=$ 5.02 TeV. The distribution in red markers shows the original event-by-event $\langle p_\mathrm{T}\rangle$ distribution in which $N_\mathrm{ch}$ fluctuates whereas the distribution shown with black markers is the $\langle p_\mathrm{T}\rangle$ distribution in which $N_\mathrm{ch}$ is fixed to $N_\mathrm{ch}^\mathrm{min}$ in each event of the centrality class. A Gaussian function is used to fit the two distributions and the ratio of data to fit is shown in the bottom panel of Fig.~\ref{fig:meanpt_dist_2}. The ratio shows that the data points are above the fit in the right-side of the distribution, and below the fit in the left-side, indicating that the $\langle p_\mathrm{T}\rangle$ distributions are positively skewed. 
The $\langle p_\mathrm{T}\rangle$ distributions exhibit substantial deviations from a Gaussian distribution even under the condition of a fixed $N_\mathrm{ch}$, thereby reflecting the analogous traits observed in the original distribution with fluctuating $N_\mathrm{ch}$. The middle and right panels of Fig.~\ref{fig:meanpt_dist_2} visually exemplify the consistent manifestation of these characteristics within the distributions of $\langle p_\mathrm{T}\rangle$ for both semicentral and peripheral collisions. It is concluded that the positive skewness in the event-by-event $\langle p_\mathrm{T}\rangle$ distributions is not a trivial consequence of event-by-event statistical fluctuations of $N_\mathrm{ch}$, and therefore new and non-trivial information can be extracted from higher-order moments of $\langle p_\mathrm{T}\rangle$ fluctuations.    

The $\langle p_\mathrm{T}\rangle$ correlators described in Eqs.~\ref{eq:simp2},~\ref{eq:simp3}, and~\ref{eq:simp4} are calculated using charged-particle tracks in different centrality (multiplicity) classes. For Pb--Pb collisions, the observables are analyzed in 18 equal-width centrality classes ranging between zero to ninety percent, i.e., 0--5\%, 5--10\%, ..., 85--90\%. For Xe--Xe collisions, the total number of events are divided into four centrality classes, 0--20\%, 20--40\%, 40--60\%, and 60--80\%. The centrality classes are taken wider for Xe--Xe collisions because  of the limited data sample. In pp collisions, the events are classified into ten multiplicity classes, 0--1\%, 1--5\%, 5--10\%, 10--15\%, 15--20\%, 20--30\%, 30--40\%, 40--50\%, 50--70\%, and 70--100\%. The method of defining these multiplicity classes can be found in Ref.~\cite{ALICE:2020swj}. 

\begin{table}
  \begin{center}
      \caption{ \label{tab:skewandkurt} Contributions to the total systematic uncertainty on standardized skewness $\gamma_{\langle p_\mathrm{T}\rangle}$, intensive skewness $\Gamma_{\langle p_\mathrm{T}\rangle}$, and kurtosis $\kappa_{\langle p_\mathrm{T}\rangle}$ in Pb--Pb collisions at $\sqrt{s_\mathrm{NN}}$ $=$ 5.02 TeV, Xe--Xe collisions at $\sqrt{s_\mathrm{NN}}$ = 5.44 TeV, and pp collisions at $\sqrt{s}$ $=$ 5.02 TeV. Ranges are given where the uncertainties depend on centrality or multiplicity.}

      \begin{tabular}{llccc}\\ \hline
      Observables & Sources of systematic uncertainty & Pb--Pb & Xe--Xe & pp \\ \hline
      & Vertex $z$-position & 0.7--2.7\%  & 1.5--9.1\% & 0.3--0.8\% \\
      & Pileup & 0.5--6.4\% & -- & --\\
      & $\chi^{2}_\mathrm{TPC}/n_\mathrm{TPCclusters}$ & 0.5--2.6\% & 3.4-6.5\% & 0.1--0.4\%\\
      & $\chi^{2}_\mathrm{ITS}/n_\mathrm{ITSclusters}$ & 0.5--2.2\% & 3.6--11.3\% & $<0.1\%$\\
      $\gamma_{\langle p_\mathrm{T}\rangle}$ & $n_{\mathrm{TPCcrossedrows}}$ & 0.6--3.8\% & 2.6--6.8\% & 0.9--1.9\%\\
      & $\mathrm{DCA}_{xy}$ & 0.6--4.5\% & 2.5--7.3\%  & 0.5--1.1\%\\
      & $\mathrm{DCA}_{z}$ & 0.4--1.8\% & 2.8--5.9\%  & $<0.1\%$ \\
      & MC closure & 2.3\% & 2.6\% & 2.7\%\\
      \cline{2-5}
      & Total & 2.9--8.7\% & 9.1--17.2\% & 2.9--3.6\%\\   
      \hline

      & Vertex $z$-position & 0.6--2.7\%  & 1.7--9.1\% & 0.3--0.8\% \\
      & Pileup & 0.5--6.3\% & -- & --\\ 
      & $\chi^{2}_\mathrm{TPC}/n_\mathrm{TPCclusters}$ & 0.4--2.7\% & 3.6--6.9\% & 0.1--0.3\%\\
      & $\chi^{2}_\mathrm{ITS}/n_\mathrm{ITSclusters}$ & 0.4--2.2\% & 3.7--11.4\% & $<0.1\%$\\
      $\Gamma_{\langle p_\mathrm{T}\rangle}$ & $n_{\mathrm{TPCcrossedrows}}$ & 0.8--3.8\% & 3.0--7.0\% & 0.3--1.7\%\\
      & $\mathrm{DCA}_{xy}$ & 0.7--4.7\% & 0.4--4.2\%  & 0.3--1\%\\
      & $\mathrm{DCA}_{z}$ & 0.4--1.9\% & 2.7--6.0\%  & $<0.1\%$ \\
      & MC closure & 3.6\% & 2.3\% & 0.4\%\\
      \cline{2-5}
      & Total & 4--9\% & 9.5--17.6\% & 0.8--2.1\%\\   
      \hline
      
      & Vertex $z$-position & 0.1--3.6\%  & 0.9--4.5\% & 0.1--1.4\% \\
      & Pileup & 0.2--4.5\% & -- & --\\ 
      & $\chi^{2}_\mathrm{TPC}/n_\mathrm{TPCclusters}$ & 0.1--1.8\% & 0.9--4.6\% & 0.1--1.1\%\\
      & $\chi^{2}_\mathrm{ITS}/n_\mathrm{ITSclusters}$ & 0.1--0.9\% & 1.2--3.9\% & $<0.1\%$\\
      $\kappa_{\langle p_\mathrm{T}\rangle}$ & $n_{\mathrm{TPCcrossedrows}}$ & 0.1--1.3\% & 3.0--7.0\% & 0.3--1.7\%\\
      & $\mathrm{DCA}_{xy}$ & 0.1--2.3\% & 0.3--0.9\%  & 0.5--3.3\%\\
      & $\mathrm{DCA}_{z}$ & 0.1--2.9\% & 0.7--1.9\%  & $<0.1\%$ \\
      & MC closure & 0.3\% & 1.3\% & 0.9\%\\
      \cline{2-5}
      & Total & 0.5--7.3\% & 2.7--11.7\% \% & 1.8--4.7\%\\   
      \hline
      \end{tabular}
  \end{center}
\end{table}

The $\langle p_\mathrm{T}\rangle$ correlators are calculated in the unit multiplicity bins of a given centrality class, and then merged using centrality bin width correction formula~\cite{Luo:2013bmi}. In this way, both the effects of multiplicity fluctuations, and the necessity of using non-trivial multiplicity weight to determine all-event averages from single-event averages, are eliminated. 
The statistical uncertainties are evaluated using the bootstrap method. By repeatedly resampling the dataset and analyzing each resample, the bootstrap method provides a robust and computationally feasible approach for estimating the uncertainties associated with the observed results. It allows for a comprehensive assessment of the variability in the data, taking into account the inherent fluctuations and dependencies present in the measurements. The systematic uncertainties on the observables are estimated separately for each collision system by varying event and track selection criteria. The uncertainties related to the event selection include the variation of the accepted vertex position along the beam direction and pileup cut where applicable. The uncertainties due to track selection include the variation of the selection criteria on $\mathrm{DCA}_{xy}$, $\mathrm{DCA}_{z}$, the number of reconstructed space points in the TPC, and the quality of the track fit from their nominal values. Systematic uncertainties obtained from the contribution of each of these sources are considered as uncorrelated and the total systematic uncertainty on the observables is obtained by adding them in quadrature. Table~\ref{tab:skewandkurt} shows the summary of the contributions to the total systematic uncertainty on standardized skewness, intensive skewness, and kurtosis in Pb--Pb, Xe--Xe, and pp collisions.

The standardized skewness, intensive skewness, and kurtosis reported in this draft are considered robust and independent of detection efficiencies~\cite{SkewnessTheoryPRC}. The detector inefficiencies cancel to leading order within these ratios, despite their potential presence in individual $p_\mathrm{T}$ correlators. Consequently, efficiency correction is not carried out for these ratios. The robustness of the measured ratios is further affirmed by performing a Monte Carlo (MC) closure test. The MC closure test uses simulations based on different event generators for producing particles that corresponds to generated (true) results and GEANT3~\cite{Brun:1994aa} for the transport of particles through the geometry of ALICE detectors. For Pb--Pb and Xe--Xe collisions, the MC events are generated using HIJING (Heavy-Ion Jet Interaction Generator)~\cite{Wang:1991hta} whereas for pp collisions, the events are produced by PYTHIA8~\cite{Sjostrand:2006za, Sjostrand:2014zea} event generator with Monash2013 tune. The experimental conditions prevailing during the data taking are accounted for in the generated events using GEANT3 by reproducing the actual configuration of different detectors during the runs. Results obtained from the generated events are compared with their corresponding reconstructed ones (without applying efficiency corrections), and they show a good agreement within uncertainties. The ratio of the reconstructed to the generated results for the observables are fitted with zeroth-order polynomial functions to quantify the amount of closure obtained. The percentages of agreement for standardized skewness, intensive skewness, and kurtosis are 97.3\%, 99.6\% and 99.1\%, respectively, in pp collisions at $\sqrt{s}$ $=$ 5.02 TeV. Similar study is also performed for the other two collision systems, Pb--Pb collisions at $\sqrt{s_\mathrm{NN}}$ $=$ 5.02 and Xe--Xe collisions at $\sqrt{s_\mathrm{NN}}$ $=$ 5.44 TeV. The percentages of closure for the  observables (standardized skewness, intensive skewness, kurtosis) in Pb--Pb (Xe--Xe) collisions are found to be 97.7\% (97.4\%), 96.4\% (97.7\%), and 99.7\% (98.7\%). These differences are added to the systematic uncertainties and reported in Table~\ref{tab:skewandkurt}.

To study how the skewness and kurtosis of $\langle p_\mathrm{T}\rangle$ distributions vary with the size of the collision system and to have a comparison between heavy-ion collisions (Pb--Pb and Xe--Xe) and small systems like pp collisions, expressing the centrality (multiplicity for pp collisions) classes as average charged particle multiplicity density ($\langle \mathrm{d}N_\mathrm{ch}/\mathrm{d}\eta\rangle_{|\eta|<0.5}$) is more suitable. The conversion of centrality to $\langle \mathrm{d}N_\mathrm{ch}/\mathrm{d}\eta\rangle_{|\eta|<0.5}$ is performed using the measured values of $\langle \mathrm{d}N_\mathrm{ch}/\mathrm{d}\eta\rangle_{|\eta|<0.5}$ for certain centrality classes from Ref.~\cite{ALICE:2015juo} for Pb--Pb collisions. It should be noted that the centrality classes used in this analysis are narrower and different from that in the reference. A linear relation is observed between average number of charged particles ($\langle N_\mathrm{acc}\rangle$) in the kinematic range ($|\eta| < 0.8$ and $0.2 < p_\mathrm{T} < 0.3$ GeV/$\it{c}$) of this analysis and $\langle \mathrm{d}N_\mathrm{ch}/\mathrm{d}\eta\rangle_{|\eta|<0.5}$ over the full centrality range. The extracted fit parameters from the linear relation between $\langle N_\mathrm{acc}\rangle$ and  $\langle \mathrm{d}N_\mathrm{ch}/\mathrm{d}\eta\rangle_{|\eta|<0.5}$ were used to assign a value of $\langle \mathrm{d}N_\mathrm{ch}/\mathrm{d}\eta\rangle_{|\eta|<0.5}$ for any value of $\langle N_\mathrm{acc}\rangle$. Hence, by calculating the $\langle N_\mathrm{acc}\rangle$ for the centrality classes of this analysis, the corresponding value of $\langle \mathrm{d}N_\mathrm{ch}/\mathrm{d}\eta\rangle_{|\eta|<0.5}$ is obtained. This method was previously used in the study of second-order fluctuations of $\langle p_\mathrm{T}\rangle$ in Ref.~\cite{ALICE:2014gvd}. Similar approach is followed for Xe--Xe collisions using $\langle \mathrm{d}N_\mathrm{ch}/\mathrm{d}\eta\rangle_{|\eta|<0.5}$ values of centrality classes from Ref.~\cite{ALICE:2018cpu}. For pp collisions, the measured values of $\langle \mathrm{d}N_\mathrm{ch}/\mathrm{d}\eta\rangle_{|\eta|<0.5}$ are taken from Ref.~\cite{ALICE:2020swj} since the multiplicity classes are the same.

\section{Results}
\label{sec:res}
\subsection{Standardized skewness}
\begin{figure}[tb]
    \begin{center}
    \includegraphics[width = 0.65\textwidth]{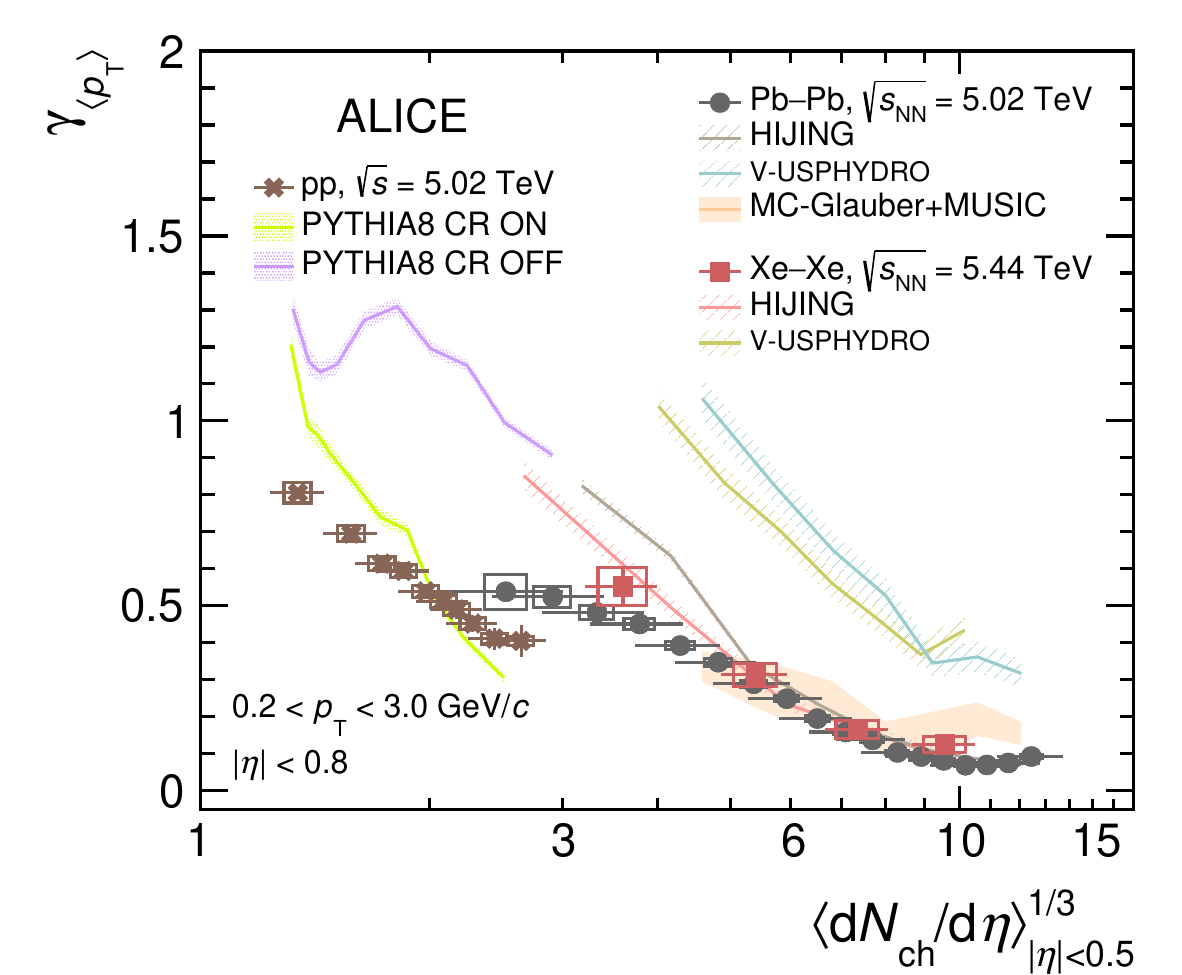}
    \end{center}
    \caption{Standardized skewness $\gamma_{\langle p_\mathrm{T}\rangle}$ shown as a function of $\langle \mathrm{d}N_\mathrm{ch}/\mathrm{d}\eta\rangle_{|\eta|<0.5}^{1/3}$ in Pb--Pb collisions at $\sqrt{s_\mathrm{NN}}$ $=$ 5.02 TeV, Xe--Xe collisions at $\sqrt{s_\mathrm{NN}}$ $=$ 5.44 TeV, and pp collisions at $\sqrt{s}$ $=$ 5.02 TeV. The predictions from different event generators and hydrodynamic model calculations from \textsc{v-usphydro}~\cite{SkewnessTheoryPRC} and MC-Glauber+MUSIC~\cite{Alver:2008aq, Schenke:2010nt} are denoted by colored bands. The statistical (systematic) uncertainties are represented by vertical bars (boxes). }
    \label{fig:standskew_hydro}
\end{figure}

The standardized skewness of charged particles produced in Pb--Pb collisions at $\sqrt{s_\mathrm{NN}}=5.02$ TeV and Xe--Xe collisions at $\sqrt{s_\mathrm{NN}}=5.44$ TeV as a function of system size denoted by the cubic root of average charged particle multiplicity density, $\langle \mathrm{d}N_\mathrm{ch}/\mathrm{d}\eta\rangle_{|\eta|<0.5}^{1/3}$ is shown in Fig.~\ref{fig:standskew_hydro}. The femtoscopic radii associated with the radius of the fireball at freeze-out, scale linearly with $\langle \mathrm{d}N_\mathrm{ch}/\mathrm{d}\eta\rangle_{|\eta|<0.5}^{1/3}$ and therefore it is used as a proxy for the system size~\cite{ALICE:2011dyt}. The uncertainties on $\langle \mathrm{d}N_\mathrm{ch}/\mathrm{d}\eta\rangle_{|\eta|<0.5}^{1/3}$ are obtained by propagating the uncertainty on $\langle \mathrm{d}N_\mathrm{ch}/\mathrm{d}\eta\rangle_{|\eta|<0.5}$. It is observed that $\gamma_{\langle p_\mathrm{T}\rangle}$ decreases with increasing $\langle \mathrm{d}N_\mathrm{ch}/\mathrm{d}\eta\rangle_{|\eta|<0.5}^{1/3}$ for both Pb--Pb and Xe--Xe collisions. This is generally expected from fluctuations in a dilution scenario caused by superposition of partially independent particle-emitting sources~\cite{STAR:2005vxr, ALICE:2014gvd}. If the fluctuations in $\langle p_\mathrm{T}\rangle$ have been purely statistical and generated by finite multiplicity ($N$), then $\gamma_{\langle p_\mathrm{T}\rangle}$ would have a dependence of $1/\sqrt{N}$. 
A slight enhancement of $\gamma_{\langle p_\mathrm{T}\rangle}$ in the most central ($\langle \mathrm{d}N_\mathrm{ch}/\mathrm{d}\eta\rangle_{|\eta|<0.5}^{1/3} > 11$) Pb--Pb collisions is also observed within uncertainty. One of the probable reasons could be the reduction of two-particle correlations $\langle \Delta p_i\Delta p_j\rangle$ towards the central collisions reported previously in Ref.~\cite{ALICE:2014gvd}. Hydrodynamic model calculations from Ref.~\cite{SkewnessTheoryPRC} that uses $\rm T_{R}ENTo$ initial conditions evolved by hydrodynamic code \textsc{v-usphydro} referred as \textsc{v-usphydro} model in the figures are compared with the data for both Pb--Pb and Xe--Xe collisions. A small specific shear viscosity, $\eta/s$ $=$ 0.047 is used in the \textsc{v-usphydro} model. As shown in Fig.~\ref{fig:standskew_hydro}, the \textsc{v-usphydro} model calculations capture the general system size dependence of the measurement, but fail to describe it quantitatively. In the model, $\gamma_{\langle p_\mathrm{T}\rangle}$ is larger in Pb--Pb collisions compared to Xe--Xe collisions for almost all multiplicities, however data do not show such system dependence. Model calculations using HIJING~\cite{Wang:1991hta}, which incorporates various phenomena like multiple minijet production with initial and final state radiation or nuclear effects such as parton shadowing and jet quenching, are shown in the figure. HIJING considers nucleus--nucleus collisions as superposition of independent binary collisions of wounded nucleons. The latest version of the model, $\text{HIJING}/B\overline{B}$ v2.0 with shadowing and jet quenching effects~\cite{ToporPop:2004lve} is used. The $\gamma_{\langle p_\mathrm{T}\rangle}$ in HIJING model exhibits a strong dependence with $\langle \mathrm{d}N_\mathrm{ch}/\mathrm{d}\eta\rangle_{|\eta|<0.5}^{1/3}$ and reproduces the measurement in the semiperipheral to semicentral region. 
Calculations performed using the MC-Glauber+MUSIC model, which adopts the Glauber Monte Carlo~\cite{Alver:2008aq} approach to generate initial conditions for the collisions, followed by the MUSIC hydrodynamic model~\cite{Schenke:2010nt} with  $\eta/s$ $=$ 0.1 are presented. Notably, both the HIJING and MC-Glauber+MUSIC model utilize the Monte Carlo Glauber initial conditions, with the primary difference lying in the subsequent evolution. While HIJING lacks the implementation of collective phenomena, MUSIC incorporates such effects. Interestingly, the results obtained for the $\gamma_{\langle p_\mathrm{T}\rangle}$ in Pb--Pb collisions remain same for both models from semiperipheral to semicentral collisions. This prompts the question of whether the $\gamma_{\langle p_\mathrm{T}\rangle}$ is primarily sensitive to the details of the initial conditions. 

The results in Pb--Pb and Xe--Xe are also compared with the measurements carried out in pp collisions at $\sqrt{s}$ $=$ 5.02 TeV. The  $\gamma_{\langle p_\mathrm{T}\rangle}$ in pp collisions shows a steeper decrease with increasing $\langle \mathrm{d}N_\mathrm{ch}/\mathrm{d}\eta\rangle_{|\eta|<0.5}^{1/3}$ compared to heavy-ion collisions. The observed $\gamma_{\langle p_\mathrm{T}\rangle}$ is also found smaller in pp collisions than in Pb--Pb collisions in the overlapping $\langle \mathrm{d}N_\mathrm{ch}/\mathrm{d}\eta\rangle_{|\eta|<0.5}^{1/3}$ region. The measurements in pp collisions are compared with PYTHIA8~\cite{Sjostrand:2014zea} model, which is a pQCD based MC event generator. The PYTHIA8 model is found to describe numerous experimental results in pp collisions at LHC energies quite successfully~\cite{ALICE:2015olq, ALICE:2017pcy, ALICE:2013rdo}. This model can involve color reconnection (CR) mechanism~\cite{Sjostrand:1987su} in pp collisions, which could contribute to explaining certain collective effects that have been observed in these collisions. 
Comparison of the measurements in pp collisions with PYTHIA8 model for both cases, with and without enabling CR mechanism (referred as PYTHIA8 CR ON and PYTHIA8 CR OFF), are presented. PYTHIA8 CR OFF model calculations utterly fail to describe the measurements both qualitatively and quantitatively. On the other hand, PYTHIA CR ON calculations reproduce the dependence of $\gamma_{\langle p_\mathrm{T}\rangle}$ with $\langle \mathrm{d}N_\mathrm{ch}/\mathrm{d}\eta\rangle_{|\eta|<0.5}^{1/3}$ although with a larger slope.

\subsection{Intensive skewness}
\begin{figure}[tb]
    \begin{center}
    \includegraphics[width = 0.65\textwidth]{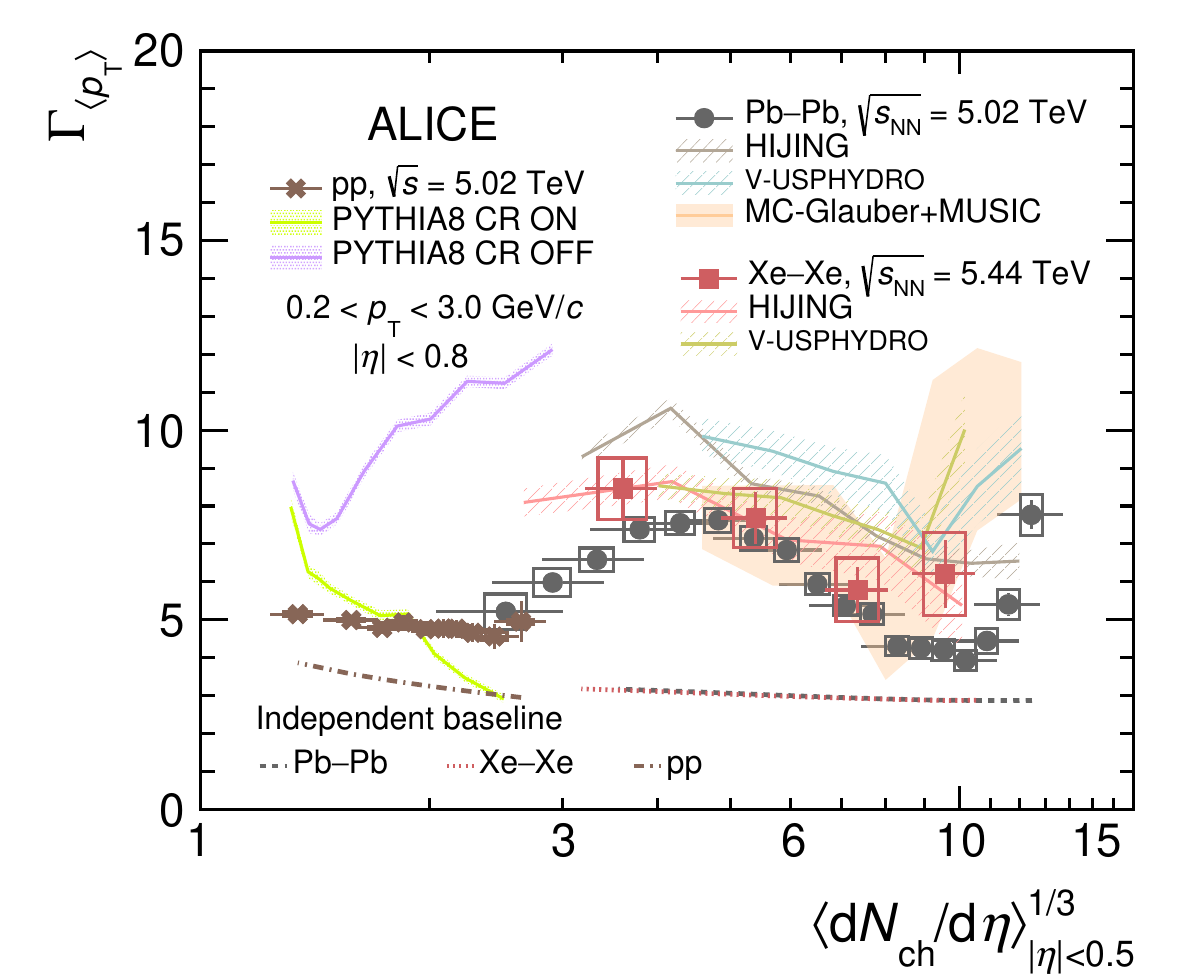}
    \end{center}
    \caption{Intensive skewness $\Gamma_{\langle p_\mathrm{T}\rangle}$ shown as a function of $\langle \mathrm{d}N_{\mathrm{ch}}/\mathrm{d}\eta\rangle_{|\eta|<0.5}^{1/3}$ in Pb--Pb collisions at $\sqrt{s_\mathrm{NN}}$ $=$ 5.02 TeV, Xe--Xe collisions at $\sqrt{s_\mathrm{NN}}$ $=$ 5.44 TeV, and pp collisions at $\sqrt{s}$ $=$ 5.02 TeV. The colored dashed lines represent the independent baseline for each system. The predictions from different event generators and hydrodynamic model calculations from \textsc{v-usphydro}~\cite{SkewnessTheoryPRC} and MC-Glauber+MUSIC~\cite{Alver:2008aq, Schenke:2010nt} are denoted by colored bands. The statistical (systematic) uncertainties are represented by vertical bars (boxes).} 
    \label{fig:intenseskew_1}
\end{figure}

Figure~\ref{fig:intenseskew_1} shows the intensive skewness studied with respect to $\langle \mathrm{d}N_\mathrm{ch}/\mathrm{d}\eta\rangle_{|\eta|<0.5}^{1/3}$ in Pb--Pb collisions at $\sqrt{s_\mathrm{NN}}$ $=$ 5.02 TeV and Xe--Xe collisions at $\sqrt{s_\mathrm{NN}}$ $=$ 5.44 TeV. The colored dashed lines indicate the independent baselines $\Gamma_\mathrm{independent}$ evaluated for each system separately. The formula~\cite{SkewnessTheoryPRC} used to calculate the independent baseline for intensive skewness is
\begin{equation}
\Gamma_\mathrm{independent} = \frac{\langle(p_\mathrm{T}-\langle p_\mathrm{T}\rangle)^{3}\rangle\langle p_\mathrm{T}\rangle}{\langle(p_\mathrm{T}-\langle p_\mathrm{T}\rangle)^{2}\rangle^2}\quad,
  \label{eq:independent}
\end{equation}
where $\langle p_\mathrm{T}\rangle$ is the average transverse momentum in given centrality class. The $p_\mathrm{T}$ spectra for different centrality classes are the input distributions for the calculation of $\Gamma_\mathrm{independent}$ as a function of centrality. The second and third central moments of the $p_\mathrm{T}$ distributions, $\langle(p_\mathrm{T}-\langle p_\mathrm{T}\rangle)^{2}\rangle$  and $\langle(p_\mathrm{T}-\langle p_\mathrm{T}\rangle)^{3}\rangle$ are evaluated using the $p_\mathrm{T}$ spectrum for each centrality class and the ratio of these moments gives the $\Gamma_\mathrm{independent}$ for that centrality class. For Pb--Pb and Xe--Xe collisions, the $p_\mathrm{T}$ spectra are taken from Refs.~\cite{ALICE:2018vuu} and~\cite{ALICE:2018hza}, respectively. Consecutively, the centralities are converted into $\langle \mathrm{d}N_\mathrm{ch}/\mathrm{d}\eta\rangle_{|\eta|<0.5}$ and $\langle \mathrm{d}N_\mathrm{ch}/\mathrm{d}\eta\rangle_{|\eta|<0.5}^{1/3}$. A positive intensive skewness, larger than the independent baseline, is observed in both Pb--Pb and Xe--Xe collisions as a function of system size. This observation is in accordance with the predictions from the \textsc{v-usphydro} hydrodynamic model calculations. The \textsc{v-usphydro} model results describe the data qualitatively but do not show quantitative agreement. The HIJING model calculations though seem to capture the decreasing trend of the measurements in the semiperipheral to semicentral region (3.8 $\leq \langle \mathrm{d}N_\mathrm{ch}/\mathrm{d}\eta\rangle_{|\eta|<0.5}^{1/3} \leq$ 8.9) but are unable to explain the rise of $\Gamma_{\langle p_\mathrm{T}\rangle}$ in the measurements from the semicentral to central region (9.0 $\leq \langle \mathrm{d}N_\mathrm{ch}/\mathrm{d}\eta\rangle_{|\eta|<0.5}^{1/3} \leq$ 12.4). The observed increase in both $\gamma_{\langle p_\mathrm{T}\rangle}$ and $\Gamma_{\langle p_\mathrm{T}\rangle}$ for $\langle \mathrm{d}N_\mathrm{ch}/\mathrm{d}\eta\rangle_{|\eta|<0.5}^{1/3} > 11$ in Pb--Pb collsions might be indicative of the onset of thermalization~\cite{Samanta:2023amp, Samanta:2023kfk}. This phenomenon is distinct from hydrodynamization, which could occur in smaller systems and at shorter timescales. As a direct consequence of thermalization, a significant increase of the skewness of $\langle p_\mathrm{T}\rangle$ fluctuations is predicted for most central collisions. If the system thermalizes, $\langle p_\mathrm{T}\rangle$ increases with density which is a characteristic phenomenon in relativistic fluid dynamics. At fixed multiplicity, larger density corresponds to smaller volume and correspondingly larger impact parameter which suggests that the rise of skewness results from a reduction in impact parameter fluctuations within the most central collisions. The MC-Glauber+MUSIC model calculations exhibit similar behavior observed in the measurements of $\Gamma_{\langle p_\mathrm{T}\rangle}$ in Pb--Pb collisions and provide better agreement quantitatively than the \textsc{v-usphydro} model results. The increase of $\Gamma_{\langle p_\mathrm{T}\rangle}$ for central collisions in the measurements, seems to be depicted by both of the hydrodynamic model calculations with different initial conditions and therfore demands theoretical inputs to understand this rise.

The measurement performed in pp collisions at $\sqrt{s}$ $=$ 5.02 TeV is also shown along with its independent baseline. For pp collisions, the baseline is calculated using the $p_\mathrm{T}$ spectra from Ref.~\cite{ALICE:2019dfi}. The hydrodynamic prediction of positive intensive skewness above of its baseline is pronounced in the measurements in pp collisions as well. 
The distinct non-monotonic behavior observed in Pb--Pb collisions as a function of $\langle \mathrm{d}N_\mathrm{ch}/\mathrm{d}\eta\rangle_{|\eta|<0.5}^{1/3}$ is noticeably attenuated in pp collisions. The $\Gamma_{\langle p_\mathrm{T}\rangle}$ exhibits a subtle, monotonic decrease as $\langle \mathrm{d}N_\mathrm{ch}/\mathrm{d}\eta\rangle_{|\eta|<0.5}^{1/3}$ increases in pp collisions. 
In the PYTHIA8 CR OFF model, the results for $\Gamma_{\langle p_\mathrm{T}\rangle}$ exhibit an increase as $\langle \mathrm{d}N_\mathrm{ch}/\mathrm{d}\eta\rangle_{|\eta|<0.5}^{1/3}$ increases, which is in direct contrast to the observations from the PYTHIA8 CR ON model. While the PYTHIA8 CR ON model appears to be in closer agreement with the experimental measurements compared to CR OFF, neither model adequately captures the qualitative behavior observed in the measurements.

\subsection{Kurtosis}
\begin{figure}[tb]
    \begin{center}
    \includegraphics[width = 0.65\textwidth]{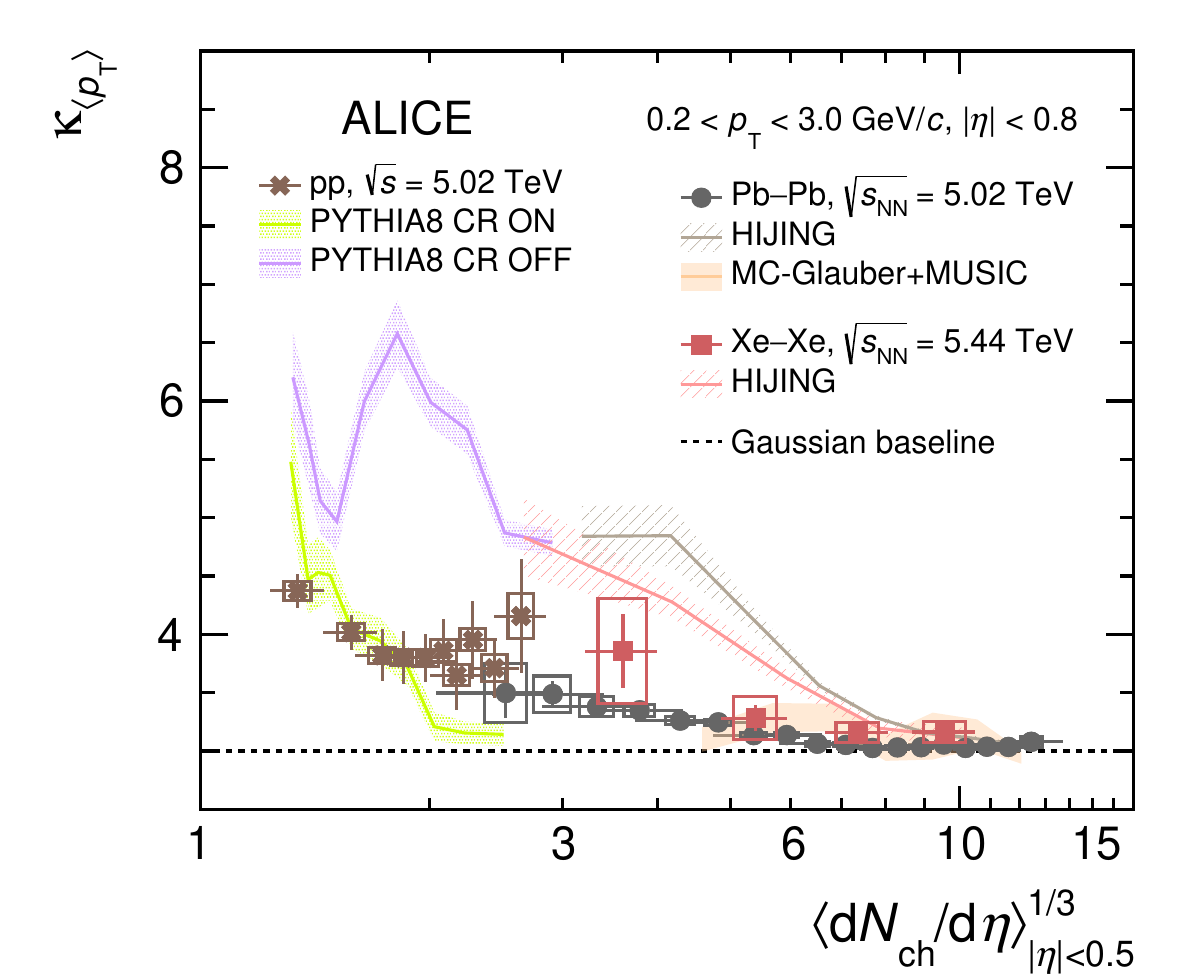}
    \end{center}
    \caption{Kurtosis $\kappa_{\langle p_\mathrm{T}\rangle}$ shown as a function of $\langle \mathrm{d}N_\mathrm{ch}/\mathrm{d}\eta\rangle_{|\eta|<0.5}^{1/3}$ in Pb--Pb collisions at $\sqrt{s_\mathrm{NN}}$ $=$ 5.02 TeV, Xe--Xe collisions at $\sqrt{s_\mathrm{NN}}$ = 5.44 TeV and pp collisions at $\sqrt{s}$ $=$ 5.02 TeV. The predictions from different event generators and hydrodynamic calculations from MC-Glauber+MUSIC model~\cite{Alver:2008aq, Schenke:2010nt} are represented by colored bands. The dashed line indicates the Gaussian baseline. The statistical (systematic) uncertainties are represented by vertical bars (boxes). }
    \label{fig:kurtosis_1}
\end{figure}

The kurtosis evaluated using Eq.~\ref{eq:simp4} as a function of $\langle \mathrm{d}N_\mathrm{ch}/\mathrm{d}\eta\rangle_{|\eta|<0.5}^{1/3}$ in Pb--Pb collisions at $\sqrt{s_\mathrm{NN}}$ $=$ 5.02 TeV and Xe--Xe collisions at $\sqrt{s_\mathrm{NN}}$ $=$ 5.44 TeV is shown in Fig.~\ref{fig:kurtosis_1}. The $\kappa_{\langle p_\mathrm{T}\rangle}$ for a Gaussian distribution, which serves as a baseline for independent particle production, is shown with dotted line. The $\kappa_{\langle p_\mathrm{T}\rangle}$ is found to decrease with increasing system size in Pb--Pb and Xe--Xe collisions and approaches the Gaussian baseline towards the most central Pb--Pb collisions. 
The measurements are compared with HIJING model calculations for both Pb--Pb and Xe--Xe collisions. While the HIJING model successfully capture the decreasing trend observed in the measurements with respect to $\langle \mathrm{d}N_\mathrm{ch}/\mathrm{d}\eta\rangle_{|\eta|<0.5}^{1/3}$, it exhibits a pronounced and rapid decrease in $\kappa_{\langle p_\mathrm{T}\rangle}$ compared to the data points. This is related to trivial system size dependence of $\kappa_{\langle p_\mathrm{T}\rangle}$ in HIJING that should follow $1/N$ dependence. Throughout the entire range of $\langle \mathrm{d}N_\mathrm{ch}/\mathrm{d}\eta\rangle_{|\eta|<0.5}^{1/3}$, the HIJING model overestimates the measurements. The MC-Glauber+MUSIC calculations reproduce the measurements in the central to midperipheral region (5.4 $\leq \langle \mathrm{d}N_\mathrm{ch}/\mathrm{d}\eta\rangle_{|\eta|<0.5}^{1/3} \leq$ 12.4). The measurements in pp collisions at $\sqrt{s}$ $=$ 5.02 TeV are also shown in the Fig.~\ref{fig:kurtosis_1}. In pp collisions, the $\kappa_{\langle p_\mathrm{T}\rangle}$ decreases with $\langle \mathrm{d}N_\mathrm{ch}/\mathrm{d}\eta\rangle_{|\eta|<0.5}^{1/3}$ and remains consistently above the Gaussian baseline even for the highest value of $\langle \mathrm{d}N_\mathrm{ch}/\mathrm{d}\eta\rangle_{|\eta|<0.5}^{1/3}$. Calculations of $\kappa_{\langle p_\mathrm{T}\rangle}$ in PYTHIA8 show similar behavior with $\langle \mathrm{d}N_\mathrm{ch}/\mathrm{d}\eta\rangle_{|\eta|<0.5}^{1/3}$ as in data when CR mechanism is turned ON. PYTHIA8 CR OFF model completely fails to describe the measured $\kappa_{\langle p_\mathrm{T}\rangle}$.    

\section{Summary}
\label{sec:summary}
In summary, first results on higher order fluctuations of mean transverse momentum ($\langle p_\mathrm{T}\rangle$) of charged particles, skewness, and kurtosis in Pb--Pb, Xe--Xe, and pp collisions at LHC energies are presented. Two measures of skewness, standardized skewness and intensive skewness, are studied using three- and two-particle $p_\mathrm{T}$ correlators as a function of system size ($\langle \mathrm{d}N_\mathrm{ch}/\mathrm{d}\eta\rangle_{|\eta|<0.5}^{1/3}$). The standardized skewness is found to decrease with increasing system size in all three collision systems. Positive intensive skewness, larger than the baseline and consistent with the predictions from the hydrodynamics studies in Ref.~\cite{SkewnessTheoryPRC}, is observed in both Pb--Pb and Xe--Xe collisions. The anticipated positive intensive skewness for A--A collisions is likewise confirmed in pp collisions at similar collision energy. HIJING model calculations, which do not incorporate hydrodynamic evolution, are able to exhibit the qualitative behavior of intensive skewness in the region of 3.8 $\leq \langle \mathrm{d}N_\mathrm{ch}/\mathrm{d}\eta\rangle_{|\eta|<0.5}^{1/3} \leq$ 8.9. The striking rise of intensive skewness in the most central region is unexplained by HIJING. This effect is, however captured in the hydrodynamic calculations of both \textsc{v-usphydro} and MC-Glauber+MUSIC models. The MC-Glauber+MUSIC model shows good agreement with the measurements of the standardized skewness and intensive skewness in the region of 4.8 $\leq \langle \mathrm{d}N_\mathrm{ch}/\mathrm{d}\eta\rangle_{|\eta|<0.5}^{1/3} \leq$ 12.4. In contrast, the \textsc{v-usphydro} model, which utilizes $\rm T_{R}ENTo$ initial conditions, overestimates both measures of skewness. The discrepancy between the results obtained from the two hydrodynamic models suggests that these measurements can provide valuable insights into the initial stages of the collision. Also, the standardized skewness in HIJING and MC-Glauber+MUSIC models exhibits similar behavior, suggesting that its sensitivity may primarily be attributed to the details of the initial conditions rather than the subsequent evolution. In pp collisions, the PYTHIA8 model with the color reconnection mechanism enabled is able to qualitatively reproduce the measurements. The kurtosis measured using the two- and four-particle $p_\mathrm{T}$ correlators shows a decrease with the system size. This decreasing behavior is reproduced by the HIJING model calculations in Pb--Pb and Xe--Xe collisions, which, however, do not show a quantitative agreement with the measurements. Hydrodynamic calculations from MC-Glauber+MUSIC model explain the measurement well in the region of 4.8 $\leq \langle \mathrm{d}N_\mathrm{ch}/\mathrm{d}\eta\rangle_{|\eta|<0.5}^{1/3} \leq$ 12.4. The value of kurtosis from the MC-Glauber+MUSIC model has a similar value to that from a Gaussian distribution. The measurements of kurtosis in Pb--Pb collisions also agreeing with the Gaussian baseline for highest values of the system size, hint about the production of a locally thermalized system in most central Pb--Pb collisions.

%%%%%%%%%%%%%%%%%%%%%%%%%%%%%%%%
% end main text 
%%%%%%%%%%%%%%%%%%%%%%%%%%%%%%%%

%%%%% acknowledgements - handled by EB chairs 
\newenvironment{acknowledgement}{\relax}{\relax}
\begin{acknowledgement}
\section*{Acknowledgements}
% add specific acknowledgements here 
% ...but please don't remove the line below: funding agencies
% will be acknowledged with a custom tex file handled by EB chairs after Collab Round 2
% Version: 2023-08-10

The ALICE Collaboration would like to thank all its engineers and technicians for their invaluable contributions to the construction of the experiment and the CERN accelerator teams for the outstanding performance of the LHC complex.
The ALICE Collaboration gratefully acknowledges the resources and support provided by all Grid centres and the Worldwide LHC Computing Grid (WLCG) collaboration.
The ALICE Collaboration acknowledges the following funding agencies for their support in building and running the ALICE detector:
A. I. Alikhanyan National Science Laboratory (Yerevan Physics Institute) Foundation (ANSL), State Committee of Science and World Federation of Scientists (WFS), Armenia;
Austrian Academy of Sciences, Austrian Science Fund (FWF): [M 2467-N36] and Nationalstiftung f\"{u}r Forschung, Technologie und Entwicklung, Austria;
Ministry of Communications and High Technologies, National Nuclear Research Center, Azerbaijan;
Conselho Nacional de Desenvolvimento Cient\'{\i}fico e Tecnol\'{o}gico (CNPq), Financiadora de Estudos e Projetos (Finep), Funda\c{c}\~{a}o de Amparo \`{a} Pesquisa do Estado de S\~{a}o Paulo (FAPESP) and Universidade Federal do Rio Grande do Sul (UFRGS), Brazil;
Bulgarian Ministry of Education and Science, within the National Roadmap for Research Infrastructures 2020-2027 (object CERN), Bulgaria;
Ministry of Education of China (MOEC) , Ministry of Science \& Technology of China (MSTC) and National Natural Science Foundation of China (NSFC), China;
Ministry of Science and Education and Croatian Science Foundation, Croatia;
Centro de Aplicaciones Tecnol\'{o}gicas y Desarrollo Nuclear (CEADEN), Cubaenerg\'{\i}a, Cuba;
Ministry of Education, Youth and Sports of the Czech Republic, Czech Republic;
The Danish Council for Independent Research | Natural Sciences, the VILLUM FONDEN and Danish National Research Foundation (DNRF), Denmark;
Helsinki Institute of Physics (HIP), Finland;
Commissariat \`{a} l'Energie Atomique (CEA) and Institut National de Physique Nucl\'{e}aire et de Physique des Particules (IN2P3) and Centre National de la Recherche Scientifique (CNRS), France;
Bundesministerium f\"{u}r Bildung und Forschung (BMBF) and GSI Helmholtzzentrum f\"{u}r Schwerionenforschung GmbH, Germany;
General Secretariat for Research and Technology, Ministry of Education, Research and Religions, Greece;
National Research, Development and Innovation Office, Hungary;
Department of Atomic Energy Government of India (DAE), Department of Science and Technology, Government of India (DST), University Grants Commission, Government of India (UGC) and Council of Scientific and Industrial Research (CSIR), India;
National Research and Innovation Agency - BRIN, Indonesia;
Istituto Nazionale di Fisica Nucleare (INFN), Italy;
Japanese Ministry of Education, Culture, Sports, Science and Technology (MEXT) and Japan Society for the Promotion of Science (JSPS) KAKENHI, Japan;
Consejo Nacional de Ciencia (CONACYT) y Tecnolog\'{i}a, through Fondo de Cooperaci\'{o}n Internacional en Ciencia y Tecnolog\'{i}a (FONCICYT) and Direcci\'{o}n General de Asuntos del Personal Academico (DGAPA), Mexico;
Nederlandse Organisatie voor Wetenschappelijk Onderzoek (NWO), Netherlands;
The Research Council of Norway, Norway;
Commission on Science and Technology for Sustainable Development in the South (COMSATS), Pakistan;
Pontificia Universidad Cat\'{o}lica del Per\'{u}, Peru;
Ministry of Education and Science, National Science Centre and WUT ID-UB, Poland;
Korea Institute of Science and Technology Information and National Research Foundation of Korea (NRF), Republic of Korea;
Ministry of Education and Scientific Research, Institute of Atomic Physics, Ministry of Research and Innovation and Institute of Atomic Physics and Universitatea Nationala de Stiinta si Tehnologie Politehnica Bucuresti, Romania;
Ministry of Education, Science, Research and Sport of the Slovak Republic, Slovakia;
National Research Foundation of South Africa, South Africa;
Swedish Research Council (VR) and Knut \& Alice Wallenberg Foundation (KAW), Sweden;
European Organization for Nuclear Research, Switzerland;
Suranaree University of Technology (SUT), National Science and Technology Development Agency (NSTDA) and National Science, Research and Innovation Fund (NSRF via PMU-B B05F650021), Thailand;
Turkish Energy, Nuclear and Mineral Research Agency (TENMAK), Turkey;
National Academy of  Sciences of Ukraine, Ukraine;
Science and Technology Facilities Council (STFC), United Kingdom;
National Science Foundation of the United States of America (NSF) and United States Department of Energy, Office of Nuclear Physics (DOE NP), United States of America.
In addition, individual groups or members have received support from:
European Research Council, Strong 2020 - Horizon 2020 (grant nos. 950692, 824093), European Union;
Academy of Finland (Center of Excellence in Quark Matter) (grant nos. 346327, 346328), Finland.

\end{acknowledgement}

%%%%%%%% Bibliography 
\bibliographystyle{utphys}   % Remember we use title in the biblio
\bibliography{bibliography}
%\input {bibliography.tex}  

%%%%%%%%%%%%%%%%%%%%%%%%%%%%%%%%
% Appendices: yours (if any) + authorlist
%%%%%%%%%%%%%%%%%%%%%%%%%%%%%%%%
\newpage
\appendix

%
%\input{} % put your appendices here (if any)
%

%%%%% Authorlist - please do not touch: handled by EB chairs 
\section{The ALICE Collaboration}
\label{app:collab}
% ALICE Collaboration author list for 2023-08-10
\begin{flushleft} 
\small

S.~Acharya\,\orcidlink{0000-0002-9213-5329}\,$^{\rm 128}$, 
D.~Adamov\'{a}\,\orcidlink{0000-0002-0504-7428}\,$^{\rm 87}$, 
G.~Aglieri Rinella\,\orcidlink{0000-0002-9611-3696}\,$^{\rm 33}$, 
M.~Agnello\,\orcidlink{0000-0002-0760-5075}\,$^{\rm 30}$, 
N.~Agrawal\,\orcidlink{0000-0003-0348-9836}\,$^{\rm 52}$, 
Z.~Ahammed\,\orcidlink{0000-0001-5241-7412}\,$^{\rm 136}$, 
S.~Ahmad\,\orcidlink{0000-0003-0497-5705}\,$^{\rm 16}$, 
S.U.~Ahn\,\orcidlink{0000-0001-8847-489X}\,$^{\rm 72}$, 
I.~Ahuja\,\orcidlink{0000-0002-4417-1392}\,$^{\rm 38}$, 
A.~Akindinov\,\orcidlink{0000-0002-7388-3022}\,$^{\rm 142}$
% ...

\section*{Affiliation Notes}

$^{\rm I}$ Deceased\\
$^{\rm II}$ Also at: Max-Planck-Institut fur Physik, Munich, Germany\\
$^{\rm III}$ Also at: Italian National Agency for New Technologies, Energy and Sustainable Economic Development (ENEA), Bologna, Italy\\
$^{\rm IV}$ Also at: Dipartimento DET del Politecnico di Torino, Turin, Italy\\
$^{\rm V}$ Also at: Department of Applied Physics, Aligarh Muslim University, Aligarh, India\\
$^{\rm VI}$ Also at: Institute of Theoretical Physics, University of Wroclaw, Poland\\
$^{\rm VII}$ Also at: An institution covered by a cooperation agreement with CERN\\

\section*{Collaboration Institutes}

$^{1}$ A.I. Alikhanyan National Science Laboratory (Yerevan Physics Institute) Foundation, Yerevan, Armenia\\
$^{2}$ AGH University of Krakow, Cracow, Poland\\
$^{3}$ Bogolyubov Institute for Theoretical Physics, National Academy of Sciences of Ukraine, Kiev, Ukraine\\
$^{4}$ Bose Institute, Department of Physics  and Centre for Astroparticle Physics and Space Science (CAPSS), Kolkata, India\\
$^{5}$ California Polytechnic State University, San Luis Obispo, California, United States\\
$^{6}$ Central China Normal University, Wuhan, China\\
$^{7}$ Centro de Aplicaciones Tecnol\'{o}gicas y Desarrollo Nuclear (CEADEN), Havana, Cuba\\
$^{8}$ Centro de Investigaci\'{o}n y de Estudios Avanzados (CINVESTAV), Mexico City and M\'{e}rida, Mexico\\
$^{9}$ Chicago State University, Chicago, Illinois, United States\\
$^{10}$ China Institute of Atomic Energy, Beijing, China\\
$^{11}$ China University of Geosciences, Wuhan, China\\
$^{12}$ Chungbuk National University, Cheongju, Republic of Korea\\
$^{13}$ Comenius University Bratislava, Faculty of Mathematics, Physics and Informatics, Bratislava, Slovak Republic\\
$^{14}$ COMSATS University Islamabad, Islamabad, Pakistan\\
$^{15}$ Creighton University, Omaha, Nebraska, United States\\
$^{16}$ Department of Physics, Aligarh Muslim University, Aligarh, India\\
$^{17}$ Department of Physics, Pusan National University, Pusan, Republic of Korea\\
$^{18}$ Department of Physics, Sejong University, Seoul, Republic of Korea\\
$^{19}$ Department of Physics, University of California, Berkeley, California, United States\\
$^{20}$ Department of Physics, University of Oslo, Oslo, Norway\\
$^{21}$ Department of Physics and Technology, University of Bergen, Bergen, Norway\\
$^{22}$ Dipartimento di Fisica, Universit\`{a} di Pavia, Pavia, Italy\\
$^{23}$ Dipartimento di Fisica dell'Universit\`{a} and Sezione INFN, Cagliari, Italy\\
$^{24}$ Dipartimento di Fisica dell'Universit\`{a} and Sezione INFN, Trieste, Italy\\
$^{25}$ Dipartimento di Fisica dell'Universit\`{a} and Sezione INFN, Turin, Italy\\
$^{26}$ Dipartimento di Fisica e Astronomia dell'Universit\`{a} and Sezione INFN, Bologna, Italy\\
$^{27}$ Dipartimento di Fisica e Astronomia dell'Universit\`{a} and Sezione INFN, Catania, Italy\\
$^{28}$ Dipartimento di Fisica e Astronomia dell'Universit\`{a} and Sezione INFN, Padova, Italy\\
$^{29}$ Dipartimento di Fisica `E.R.~Caianiello' dell'Universit\`{a} and Gruppo Collegato INFN, Salerno, Italy\\
$^{30}$ Dipartimento DISAT del Politecnico and Sezione INFN, Turin, Italy\\
$^{31}$ Dipartimento di Scienze MIFT, Universit\`{a} di Messina, Messina, Italy\\
$^{32}$ Dipartimento Interateneo di Fisica `M.~Merlin' and Sezione INFN, Bari, Italy\\
$^{33}$ European Organization for Nuclear Research (CERN), Geneva, Switzerland\\
$^{34}$ Faculty of Electrical Engineering, Mechanical Engineering and Naval Architecture, University of Split, Split, Croatia\\
$^{35}$ Faculty of Engineering and Science, Western Norway University of Applied Sciences, Bergen, Norway\\
$^{36}$ Faculty of Nuclear Sciences and Physical Engineering, Czech Technical University in Prague, Prague, Czech Republic\\
$^{37}$ Faculty of Physics, Sofia University, Sofia, Bulgaria\\
$^{38}$ Faculty of Science, P.J.~\v{S}af\'{a}rik University, Ko\v{s}ice, Slovak Republic\\
$^{39}$ Frankfurt Institute for Advanced Studies, Johann Wolfgang Goethe-Universit\"{a}t Frankfurt, Frankfurt, Germany\\
$^{40}$ Fudan University, Shanghai, China\\
$^{41}$ Gangneung-Wonju National University, Gangneung, Republic of Korea\\
$^{42}$ Gauhati University, Department of Physics, Guwahati, India\\
$^{43}$ Helmholtz-Institut f\"{u}r Strahlen- und Kernphysik, Rheinische Friedrich-Wilhelms-Universit\"{a}t Bonn, Bonn, Germany\\
$^{44}$ Helsinki Institute of Physics (HIP), Helsinki, Finland\\
$^{45}$ High Energy Physics Group,  Universidad Aut\'{o}noma de Puebla, Puebla, Mexico\\
$^{46}$ Horia Hulubei National Institute of Physics and Nuclear Engineering, Bucharest, Romania\\
$^{47}$ HUN-REN Wigner Research Centre for Physics, Budapest, Hungary\\
$^{48}$ Indian Institute of Technology Bombay (IIT), Mumbai, India\\
$^{49}$ Indian Institute of Technology Indore, Indore, India\\
$^{50}$ INFN, Laboratori Nazionali di Frascati, Frascati, Italy\\
$^{51}$ INFN, Sezione di Bari, Bari, Italy\\
$^{52}$ INFN, Sezione di Bologna, Bologna, Italy\\
$^{53}$ INFN, Sezione di Cagliari, Cagliari, Italy\\
$^{54}$ INFN, Sezione di Catania, Catania, Italy\\
$^{55}$ INFN, Sezione di Padova, Padova, Italy\\
$^{56}$ INFN, Sezione di Pavia, Pavia, Italy\\
$^{57}$ INFN, Sezione di Torino, Turin, Italy\\
$^{58}$ INFN, Sezione di Trieste, Trieste, Italy\\
$^{59}$ Inha University, Incheon, Republic of Korea\\
$^{60}$ Institute for Gravitational and Subatomic Physics (GRASP), Utrecht University/Nikhef, Utrecht, Netherlands\\
$^{61}$ Institute of Experimental Physics, Slovak Academy of Sciences, Ko\v{s}ice, Slovak Republic\\
$^{62}$ Institute of Physics, Homi Bhabha National Institute, Bhubaneswar, India\\
$^{63}$ Institute of Physics of the Czech Academy of Sciences, Prague, Czech Republic\\
$^{64}$ Institute of Space Science (ISS), Bucharest, Romania\\
$^{65}$ Institut f\"{u}r Kernphysik, Johann Wolfgang Goethe-Universit\"{a}t Frankfurt, Frankfurt, Germany\\
$^{66}$ Instituto de Ciencias Nucleares, Universidad Nacional Aut\'{o}noma de M\'{e}xico, Mexico City, Mexico\\
$^{67}$ Instituto de F\'{i}sica, Universidade Federal do Rio Grande do Sul (UFRGS), Porto Alegre, Brazil\\
$^{68}$ Instituto de F\'{\i}sica, Universidad Nacional Aut\'{o}noma de M\'{e}xico, Mexico City, Mexico\\
$^{69}$ iThemba LABS, National Research Foundation, Somerset West, South Africa\\
$^{70}$ Jeonbuk National University, Jeonju, Republic of Korea\\
$^{71}$ Johann-Wolfgang-Goethe Universit\"{a}t Frankfurt Institut f\"{u}r Informatik, Fachbereich Informatik und Mathematik, Frankfurt, Germany\\
$^{72}$ Korea Institute of Science and Technology Information, Daejeon, Republic of Korea\\
$^{73}$ KTO Karatay University, Konya, Turkey\\
$^{74}$ Laboratoire de Physique Subatomique et de Cosmologie, Universit\'{e} Grenoble-Alpes, CNRS-IN2P3, Grenoble, France\\
$^{75}$ Lawrence Berkeley National Laboratory, Berkeley, California, United States\\
$^{76}$ Lund University Department of Physics, Division of Particle Physics, Lund, Sweden\\
$^{77}$ Nagasaki Institute of Applied Science, Nagasaki, Japan\\
$^{78}$ Nara Women{'}s University (NWU), Nara, Japan\\
$^{79}$ National and Kapodistrian University of Athens, School of Science, Department of Physics , Athens, Greece\\
$^{80}$ National Centre for Nuclear Research, Warsaw, Poland\\
$^{81}$ National Institute of Science Education and Research, Homi Bhabha National Institute, Jatni, India\\
$^{82}$ National Nuclear Research Center, Baku, Azerbaijan\\
$^{83}$ National Research and Innovation Agency - BRIN, Jakarta, Indonesia\\
$^{84}$ Niels Bohr Institute, University of Copenhagen, Copenhagen, Denmark\\
$^{85}$ Nikhef, National institute for subatomic physics, Amsterdam, Netherlands\\
$^{86}$ Nuclear Physics Group, STFC Daresbury Laboratory, Daresbury, United Kingdom\\
$^{87}$ Nuclear Physics Institute of the Czech Academy of Sciences, Husinec-\v{R}e\v{z}, Czech Republic\\
$^{88}$ Oak Ridge National Laboratory, Oak Ridge, Tennessee, United States\\
$^{89}$ Ohio State University, Columbus, Ohio, United States\\
$^{90}$ Physics department, Faculty of science, University of Zagreb, Zagreb, Croatia\\
$^{91}$ Physics Department, Panjab University, Chandigarh, India\\
$^{92}$ Physics Department, University of Jammu, Jammu, India\\
$^{93}$ Physics Program and International Institute for Sustainability with Knotted Chiral Meta Matter (SKCM2), Hiroshima University, Hiroshima, Japan\\
$^{94}$ Physikalisches Institut, Eberhard-Karls-Universit\"{a}t T\"{u}bingen, T\"{u}bingen, Germany\\
$^{95}$ Physikalisches Institut, Ruprecht-Karls-Universit\"{a}t Heidelberg, Heidelberg, Germany\\
$^{96}$ Physik Department, Technische Universit\"{a}t M\"{u}nchen, Munich, Germany\\
$^{97}$ Politecnico di Bari and Sezione INFN, Bari, Italy\\
$^{98}$ Research Division and ExtreMe Matter Institute EMMI, GSI Helmholtzzentrum f\"ur Schwerionenforschung GmbH, Darmstadt, Germany\\
$^{99}$ Saga University, Saga, Japan\\
$^{100}$ Saha Institute of Nuclear Physics, Homi Bhabha National Institute, Kolkata, India\\
$^{101}$ School of Physics and Astronomy, University of Birmingham, Birmingham, United Kingdom\\
$^{102}$ Secci\'{o}n F\'{\i}sica, Departamento de Ciencias, Pontificia Universidad Cat\'{o}lica del Per\'{u}, Lima, Peru\\
$^{103}$ Stefan Meyer Institut f\"{u}r Subatomare Physik (SMI), Vienna, Austria\\
$^{104}$ SUBATECH, IMT Atlantique, Nantes Universit\'{e}, CNRS-IN2P3, Nantes, France\\
$^{105}$ Sungkyunkwan University, Suwon City, Republic of Korea\\
$^{106}$ Suranaree University of Technology, Nakhon Ratchasima, Thailand\\
$^{107}$ Technical University of Ko\v{s}ice, Ko\v{s}ice, Slovak Republic\\
$^{108}$ The Henryk Niewodniczanski Institute of Nuclear Physics, Polish Academy of Sciences, Cracow, Poland\\
$^{109}$ The University of Texas at Austin, Austin, Texas, United States\\
$^{110}$ Universidad Aut\'{o}noma de Sinaloa, Culiac\'{a}n, Mexico\\
$^{111}$ Universidade de S\~{a}o Paulo (USP), S\~{a}o Paulo, Brazil\\
$^{112}$ Universidade Estadual de Campinas (UNICAMP), Campinas, Brazil\\
$^{113}$ Universidade Federal do ABC, Santo Andre, Brazil\\
$^{114}$ Universitatea Nationala de Stiinta si Tehnologie Politehnica Bucuresti, Bucharest, Romania\\
$^{115}$ University of Cape Town, Cape Town, South Africa\\
$^{116}$ University of Derby, Derby, United Kingdom\\
$^{117}$ University of Houston, Houston, Texas, United States\\
$^{118}$ University of Jyv\"{a}skyl\"{a}, Jyv\"{a}skyl\"{a}, Finland\\
$^{119}$ University of Kansas, Lawrence, Kansas, United States\\
$^{120}$ University of Liverpool, Liverpool, United Kingdom\\
$^{121}$ University of Science and Technology of China, Hefei, China\\
$^{122}$ University of South-Eastern Norway, Kongsberg, Norway\\
$^{123}$ University of Tennessee, Knoxville, Tennessee, United States\\
$^{124}$ University of the Witwatersrand, Johannesburg, South Africa\\
$^{125}$ University of Tokyo, Tokyo, Japan\\
$^{126}$ University of Tsukuba, Tsukuba, Japan\\
$^{127}$ Universit\"{a}t M\"{u}nster, Institut f\"{u}r Kernphysik, M\"{u}nster, Germany\\
$^{128}$ Universit\'{e} Clermont Auvergne, CNRS/IN2P3, LPC, Clermont-Ferrand, France\\
$^{129}$ Universit\'{e} de Lyon, CNRS/IN2P3, Institut de Physique des 2 Infinis de Lyon, Lyon, France\\
$^{130}$ Universit\'{e} de Strasbourg, CNRS, IPHC UMR 7178, F-67000 Strasbourg, France, Strasbourg, France\\
$^{131}$ Universit\'{e} Paris-Saclay, Centre d'Etudes de Saclay (CEA), IRFU, D\'{e}partment de Physique Nucl\'{e}aire (DPhN), Saclay, France\\
$^{132}$ Universit\'{e}  Paris-Saclay, CNRS/IN2P3, IJCLab, Orsay, France\\
$^{133}$ Universit\`{a} degli Studi di Foggia, Foggia, Italy\\
$^{134}$ Universit\`{a} del Piemonte Orientale, Vercelli, Italy\\
$^{135}$ Universit\`{a} di Brescia, Brescia, Italy\\
$^{136}$ Variable Energy Cyclotron Centre, Homi Bhabha National Institute, Kolkata, India\\
$^{137}$ Warsaw University of Technology, Warsaw, Poland\\
$^{138}$ Wayne State University, Detroit, Michigan, United States\\
$^{139}$ Yale University, New Haven, Connecticut, United States\\
$^{140}$ Yonsei University, Seoul, Republic of Korea\\
$^{141}$  Zentrum  f\"{u}r Technologie und Transfer (ZTT), Worms, Germany\\
$^{142}$ Affiliated with an institute covered by a cooperation agreement with CERN\\
$^{143}$ Affiliated with an international laboratory covered by a cooperation agreement with CERN.\\

\end{flushleft} 

\end{document}